\documentclass[preprint]{aastex}

\email{STATUS: in press//
ApJ v582 n2 January 10, 2003}

\shorttitle{The Global Baroclinic Instability in Accretion Disks.}
\shortauthors{Klahr and Bodenheimer}

\begin{document}


\title{Turbulence in Accretion Disks.\\ 
Vorticity Generation and Angular Momentum Transport\\ 
via the Global Baroclinic Instability$^\dagger$}


\author{H.H.\ Klahr \altaffilmark{1}}
\email{klahr@ucolick.org}

\and

\author{P.\ Bodenheimer}

\affil{UCO/Lick Observatory, University of California,
    Santa Cruz, CA 95064}


\altaffiltext{1}{present address: Computational Physics, Institut f\"ur Astronomie und Astrophysik, Universit\"at T\"ubingen, Germany}

\newcommand{\tramp}{{\sc Tramp\,\,}}
\newcommand{\okappa}{{\tilde{\kappa}}}

\begin{abstract}
In this paper we present the global baroclinic instability as a
source for vigorous turbulence leading to angular momentum transport 
in Keplerian accretion disks.
We show by analytical considerations and three-dimensional  radiation hydro
simulations that, in particular, protoplanetary disks have a negative radial 
entropy gradient, which makes
them baroclinic. 
Two-dimensional numerical simulations show that 
a baroclinic flow is unstable and produces turbulence. These findings are 
tested for numerical effects by performing a
simulation with a barotropic initial condition which shows that imposed turbulence rapidly decays. 
The turbulence in baroclinic disks transports angular momentum outward and  creates a radially  
inward bound accretion of matter. Potential energy is released and excess kinetic energy is dissipated.
Finally the reheating of the gas supports the radial entropy gradient, forming a self consistent process.
We measure accretion rates in our 2D and 3D simulations of $\dot M= - 10^{-9}~{\rm to~}-10^{-7}~\mbox{M}_{\sun}~
 \mbox{yr}^{-1}$ 
and viscosity parameters of $\alpha = 10^{-4} - 10^{-2}$, which
fit perfectly together and agree reasonably with observations.
The turbulence creates
pressure waves, Rossby waves, and vortices in the ($R-\phi$) plane of the disk.
We demonstrate in a global simulation that these vortices tend 
to form out of little background noise and to 
be long-lasting features, which have already been suggested to lead to 
the formation of planets. 
\end{abstract}


\noindent
$^\dagger$UCO/Lick Observatory Bulletin, No..........


\section{Introduction}
Protoplanetary disks appear to be a common feature
around young stars (Beckwith \& Sargent 1993; Strom, Edwards, 
\& Skrutskie 1993). 
They are thought to provide the material and the environment
for the formation of planets (e.g.\ Lissauer 1993).
Thus one needs to know the internal properties of such
disks, such as the  density, temperature and turbulence,  in order
to estimate the time scales of the formation process. 
These quantities are not directly accessible via
observation, and so one needs a model for these disks
to derive observable quantities like line emission
and scattering efficiency for the light from the 
central star (Bell, Cassen, Klahr, \& Henning 1997; 
Hartmann, Calvet, Gullbring, \& D'Alessio 1998; Bell 1999).

   The basic idea of most models is that there
is a process in these disks that transfers angular momentum
radially outward, so that mass will flow radially inward (L\"ust 1952).
Such a process might be turbulence (hydrodynamical or
magneto-hydrodynamical) or self gravity (e.g.\ Larson 1989; Stone, 
Gammie, Balbus,  \&  Hawley 2000). Independent
of the source for the angular momentum transport, it
can be parameterized by an effective viscosity $\nu$, which
usually is scaled to  the local sound speed $c_s$ and the
pressure scale height $H_D$ of the disk by  a dimensionless
number called $\alpha$ (Shakura \& Sunyaev 1973):
\begin{equation}
\nu = \alpha c_s H_D~.
\end{equation}
Using this simple description, it was possible to calculate
the time evolution of disks (Lynden-Bell \& Pringle 1974; Lin \& Papaloizou
1980),
the density and temperature
structure, and the turbulent background as well as the laminar 
sub Keplerian mean component of the gas flow. The $\alpha$ models
have influenced our understanding of the planetary formation process
through their implications, for example, regarding the spatial distribution and collision rate of dust 
grains (Markiewicz, Mizuno,  \& Voelk 1991; Weidenschilling \& Cuzzi 1993;
 Dubrulle, Morfill,  \& Sterzik 1995; Klahr \& Henning 1997), the opening of gaps by giant planets 
 (Bryden, Chen, Lin, Nelson, \& Papaloizou 1999;
Kley 1999) or the radial drift rate 
of planets during their formation phas	e (Goldreich \& Tremaine 1980; Ward 1997).

They are also used to explain the generation 
of FU-Orionis outbursts (Bell \& Lin 1994; Kley \& Lin 1999) as well as the measured accretion rates
of T-Tauri stars (Hartmann et al.\ 1998).

   Despite the success of these models, $\alpha$  is still
a free parameter for protoplanetary accretion disks.
A barotropic Keplerian shear flow is in principle stable (Rayleigh stable) 
and does not develop turbulence per se despite the high Reynolds numbers.
Dubrulle (1993), Richard \& Zahn (1999), and Duschl, Strittmatter, \& Biermann (2000)
claim that the high 
Reynolds numbers will lead to turbulence in barotropic disks, but
numerical investigations so far contradict this idea (Balbus, Hawley,  \& Stone 1996;
Godon \& Livio 1999a).  Hydrodynamical turbulence has shown
to transport angular momentum outward (Hawley, Balbus, \& Winters 1999; R\"udiger \& Drecker 2001), but it is not able to sustain itself 
and usually rapidly decays in barotropic simulations.
For a long time thermal convection in the vertical direction was thought to provide
a turbulent transport of angular momentum in the outward direction (e.g. Cameron 1978; Lin \&
Papaloizou 1985). However analytical investigations of convection 
 indicate inward transport of angular momentum (Ryu \& Goodman 1992), a result that was
later confirmed by numerical simulations (see below). 
For disks around black holes or in cataclysmic variables, 
magneto-hydrodynamical instabilities (e.g. magneto-rotational instability)
seem to provide a reasonable amount of viscosity (Balbus \& Hawley 1991). 
Relatively massive protostellar disks around young stars  ($> 0.1 M_\star$) 
also have a proper mechanism to transport
angular momentum via self gravity (Toomre 1964; Laughlin \& Bodenheimer 1994).

   In the less massive protoplanetary accretion disks ($\approx 0.01 M_\star$)
self gravity is only important at the outermost radii (Bell et al.\ 1997).
At the same time the disk is so cool and contaminated
by tiny dust grains,  which capture most  of the few free electrons,
that magnetic fields have negligible influence on the bulk of
the disk matter in the regions where planets are expected 
to form (Gammie 1996).
The sub-micron sized
dust grains reduce the free electron density
efficiently enough to 
prevent the possibility of the magneto-rotational
instability, as the field lines can diffuse faster than
the shear can tangle the field.
Only in the vicinity of the star, where dust has evaporated
and no planets can form, is there  sufficient ionization for 
the development of  magnetic turbulence, which can generate 
 $\alpha \approx  0.01$ (Stone et al.\ 2000). 
In addition, if the optical depth above the disk is sufficiently small,
cosmic rays and X-rays (Glassgold, Najita, \& Igea 1997)
could ionize the uppermost thin surface layers of 
the disk in the hypothetical absence of dust, which has 
led some authors to the speculation that between about 0.1 and 30 AU
a disk consists of a non-accreting `dead zone'
sandwiched between
two active layers at the surface (Gammie 1996).

   Indeed, if there is no self gravity working, basically no
ionization present and no shear instability possible, the
disk would develop no turbulence, transport no angular momentum,
release no accretion energy and would basically cool down to 
the ambient temperature, while orbiting the star on a time-constant
orbit. In such a disk, time scales for the growth of planetesimals
would be much longer than in the standard scenarios, as the Brownian
motion of the micron-sized dust is smaller and there is  no
turbulence acting on the dust in the cm- to m-size range.
Sedimentation of the grains to the midplane could create a
shear instability (Cuzzi, Dobrovolskis, \& Champney 1993) which then generates
turbulence. Some authors (Cuzzi, Dobrovolskis, \& Hogan 1996) argue
that the solar nebula at one time must have been turbulent in order
to explain the size segregation effects during the formation of chondrites.
In any case we want to stress that it is of
major relevance for planet formation to know whether protoplanetary disks are
turbulent or not. 

   In this paper we present numerical simulations of a purely 
hydrodynamical instability that works
in accretion disks, namely 
a baroclinic instability, similar to the one
responsible for turbulent patterns on planets, for example, 
Jupiter's red spot and the weather patterns of cyclones 
and anti-cyclones on earth. 
   Baroclinic instabilities arise in rotating fluids when
surfaces of constant density are inclined with respect to the
surfaces of constant pressure (e.g.\ Tritton \& Davies 1985).
Vortensity, defined  as vorticity per unit surface density,
 is not conserved as is the 
case in barotropic two-dimensional flows
(Kelvin's circulation theorem, see e.g. Pedlosky 1987), and 
vortices can be generated.

Kippenhahn \& Thomas (1982) studied 
the compatibility of thermal and hydrostatic equilibrium in thin radiative accretion
disks but only checked axisymmetric flows. Cabot (1984) investigated the possibility and efficiency
of a baroclinic instability for cataclysmic variables (CV) in a local 
fashion, concentrating on local vertical fluctuations. He found
that this instability  produces insufficient viscosity to explain CVs, 
but nevertheless enough viscosity ($\alpha \approx 10^{-2}$)
 for protoplanetary disks.
Knobloch \& Spruit (1986) also investigated the baroclinic instability
due to the vertical stratification in the disk but found that
it is not a reliable mechanism for angular momentum transport
in thin disks. 
Ryu \& Goodman (1992) considered linear growth of non-axisymmetric
disturbances in convectively unstable disks. They used the shearing-sheet
approximation in a uniform disk and found that the flux of angular
momentum was inwards.  Lin, Papaloizou, \& Kley    (1993)
performed a linear stability analysis of 
non-axisymmetric convective instabilities
in disks but allowed for some disk 
structure in the radial direction. Their
disk model includes a small radial interval in which the 
entropy has a small local maximum of 7$\%$ above the background
but with a steep drop with an average slope of $K\sim R^{-3}$ ($K$ is the 
polytropic constant; see below). 
Such a situation could be
baroclinically unstable, and in
fact that region is found to be associated with outward transport of
angular momentum. 
Lovelace, Li, Colgate, \& Nelson (1999) and Li, Finn, Lovelace, \& Colgate (2000)
investigated
the stability of a strong local entropy maximum (a factor 3 above the 
background with a slope  
$K \sim R^{-11.5}$) 
in a thin Keplerian
disk and found the situation 
to be unstable to the formation of Rossby waves, which transported
angular momentum outward and ultimately formed vortices (Li, Colgate, Wendroff, Liska 2001).
Sheehan, Davies, Cuzzi, \& Estberg (1999) studied the
propagation and generation of Rossby waves in the
protoplanetary nebula in great detail, but they had to 
assume some turbulence as a prerequisite. 

   In contrast, here we investigate a baroclinic instability
that arises from the general
 radial (global) stratification of the gas flow in accretion disks.
Please keep in mind that global refers to baroclinic and does not necessarily
imply that the instability has a global character, i.e.\ depends on
the boundary conditions.
Our motivation is based on the observation of positive
Reynolds stresses in radiation-hydrodynamical 
three-dimensional (3D) simulations of thermal convection 
in protoplanetary accretion disks (Klahr \& Bodenheimer 2000a). 
As convection is known to have  the property of transporting 
angular momentum inward (Kley, Papaloizou, \& Lin 1993; Cabot 1996, 
hereafter referred to as C96;  Stone \& Balbus 1996, hereafter referred to 
as SB96),
we were surprised by this result and started an investigation to identify 
the special ingredient that would explain this contradictory      
result. We found that the size of the simulation domain, especially the 
azimuthal extent, influences the sign of the Reynolds stresses 
(Klahr \& Bodenheimer 2000a,b). 
In order to accomplish a sufficient number of tests to isolate  the 
crucial ingredient in our model and to show that artificial boundary effects
are not contributing, 
we stripped the 3D radiation-hydro simulation
down to a flat 2D ($r-\phi$) disk calculation. The 
radially varying initial temperature ($\approx$ Entropy $\sim K$) in these calculations approximately reproduces
the temperature distribution found in the 3D simulations. These tests
show that indeed there is angular momentum transport outwards and that
the lower-order azimuthal modes give the fastest growth rates in baroclinic
simulations.

In \S 2 we make an argument to show 
that protoplanetary accretion disks can be unstable
to a non-axisymmetric global baroclinic instability as long as the source term
of vorticity (baroclinic term) does not vanish. In \S  3 we explain
the changes to the TRAMP code (Klahr, Henning, \& Kley  1999)
that became necessary for the  simulations presented here. 
Two results from 3D radiation hydro simulations are shown in \S 4.
One model 
adopts an artificial heating term similar to that in 
C96 and SB96, while the other one is self-consistent in the
sense that the only possible heating is a pure effect of compression and
shock dissipation.
In \S 5 we describe tests with the 2D flat 
approximation which prove numerically the existence of the 
instability and show the strength of the turbulence that 
develops as well as the Reynolds stresses that are produced.
An initially isothermal (= globally barotropic) calculation which shows decaying turbulence
demonstrates the reliability of our numerical investigation
and proves that the instabilities observed in baroclinic models
are not an effect of the shearing disk boundary conditions, or our numerical scheme.
In \S 6 we discuss the first global (2D)  simulation of
an accretion disk allowing for the baroclinic instability
and find the interesting result that large vortices form. 
These vortices are long-lived high-pressure 
anti-cyclones with an over-density by a factor of up to four.
It has been suggested that such vortices could be 
 direct precursors of proto-planets, whose formation could
be initiated by concentration of dust toward their centers
(Barge \& Sommeria 1995; Tanga, Babiano, Dubrulle, \&  Provenzale 1996; Godon \& Livio 1999b).
There is also the possibility  that the over-dense regions could eventually
undergo
gravitational collapse (see Adams \& Watkins 1995), but our 
present code does not allow for this process.
In the last section {(\S 7)} we discuss our results.

\section{Stability considerations}
In this section we show that a baroclinic instability is plausible in a disk. 
The instability can only arise if there is an inclination
between the density and 
pressure gradients (baroclinic term): 
\begin{equation}
\nabla \rho \times \nabla p \neq 0.
\label{baro_term.ref}
\end{equation}
Here we analyze the situation using the standard polytropic equation  
 
\begin{equation}
p = K \rho^{\gamma} 
\end{equation}
where $p$ is the pressure and $\rho$ is the mass density. 
We choose a value of $\gamma = 1.43$ which is representative for a mixture
of molecular hydrogen and neutral helium with
typical abundances for protoplanetary accretion disks.
If $K$ is constant as a function of $R$, 
as used by various authors (e.g.\ Adams \& Watkins 1995,
1996;
 Goldreich, Goodman,  \& Narayan 1986; Nautha \& Toth 1998;
Godon \& Livio 1999a,b),
the baroclinic term vanishes and vorticity is conserved in plane parallel flows.

In contrast to previous numerical work we
choose a radially  varying entropy $K(R)$ (\S 5.2) for the initial state of the 2D simulations.
This is the best way to mimic the density
and temperature
distribution that arises in the 3D radiation-hydro calculation.
In other words any realistic disk has a radial entropy ($S$) gradient
($p/{\rho^{\gamma}} \neq {\rm const}$).
This can be seen from a simple argument, where we only assume that
the aspect ratio of the disk ($H_d/R$) is constant for a range of radii. 
The pressure is never only a function of local density but is also
a function of local gravity, both of which change with radius: 
\begin{equation}
p \sim \Sigma H_d \Omega^2 \sim \rho \Omega^2 R^2 \sim \rho R^{-1} 
\sim  \Sigma R^{-2}.
\end{equation}
This result holds for any accretion disk, as long as $H_d/R$ is constant. 
The exact dependence of $K$ on $R$
 nevertheless depends on the specifics of a given simulation.
For example in a thermally convective region a density distribution of
$\rho \propto  R^{-1}$ and a 
temperature distribution of  $T\propto  R^{-1}$ are  typical. Thus it follows
\begin{equation}
p \propto  T \rho \propto  K(R) \rho^{\gamma} \,\,\, => \,\,\,
 K(R) \propto  R^{-2 + \gamma}.
\end{equation}

In a more general fashion, if  $\rho \propto  R^{-\beta_\rho}$,
$T \propto  R^{-\beta_T}$, and $K \propto  R^{-\beta_K}$ this equation reads:
\begin{equation}
\beta_K = \beta_T + \beta_\rho \left(1 - \gamma\right).
\label{beta_K.ref}
\end{equation}
It obvious, that there are certain stable
profiles. For if $\beta_T = \beta_\rho \left(\gamma - 1\right)$
then $\beta_K = 0$ (e.g.\ for $\beta_{\rho} = 1 => \beta_T = 0.43$).
But our 3D radiation-hydro calculations do not show
this particular  profile (see \S \ 4).
Thus  $\beta_K = 0.57$ for the initial state in our baroclinic simulations. 

It follows from equations (\ref{baro_term.ref}) and (\ref{beta_K.ref})
that 
\begin{equation}
\nabla \rho \times \nabla (K(R) \rho ^\gamma) \sim -  \beta_K \left(\partial_\phi \rho\right), 
\label{eq:unstable} 
\end{equation}
which is always non-zero if there is an azimuthal density fluctuation and 
a non-zero $\beta_K$.
On the other hand, isothermal disks are always barotropic by definition.

Baroclinic instabilities have been widely studied in the context
of meteorology and oceanography (e.g.\ Tritton \& Davies 1985, Pedlosky 1987). There are some theoretical models
as well as some laboratory experiments which can help to understand
the basic mechanism of this instability (Sakai, Iizawa, \& Aramaki 1997).
The onset of a baroclinic instability lies in the radial
entropy gradient. This entropy gradient can result from the 
temperature gradient between the equator and north pole of
the earth.
It can also be realized in an experiment in which a fluid is
contained between two concentric cylinders - one is heated and the other one is cooled down.
In the absence of rotation
thermal convection will occur in the radial direction.
This would lead to one large scale convection cell between
equator and north pole (Hadley cell, Hadley 1735) as well as to thermal
convection between the cylinders.
But the rotation of the earth (or the rotation of the cylinders) 
inhibits this convection as it allows only
for concentric flows and no convective heat 
exchange occurs in the radial direction.
But if the entropy gradient is strong enough, 
the concentric flow becomes unstable and begins to meander. 
This is the baroclinic instability. 
As the water meanders, it is once again able to 
transport heat radially.
The same happens in the earth's  atmosphere, where the baroclinic
instability leads to meteorological effects at mid-latitudes
like the meandering jet stream and the high and low pressure systems 
(anti-cyclones and cyclones),  determining our daily weather.

   Protoplanetary accretion disks are believed to have thermal convection
perpendicular to their midplane (Cameron 1978; Lin \&
Papaloizou 1985). The disks do not fulfill the Schwarzschild
criterion for vertical stability at a wide range of radii. 
Also for the radial stratification 
a Schwarzschild criterion can be formulated
which indicates that if rotational effects were
negligible there would be radial thermal
convection occurring (e.g.\ convective zones in stars).

The Schwarzschild criterion for stability of the radial stratification 
is actually never fulfilled in our baroclinic disks,  as the radial 
component of the adiabatic temperature gradient (see Kippenhahn
\& Weigert 1990) is
\begin{equation}
\nabla_{ad} = \frac{P}{T}\frac{\partial T}{\partial P} = 1- \frac{1}{\gamma} = 0.3
\end{equation}
while the absolute temperature gradient in the radial direction is
\begin{equation}
\nabla = \frac{P}{T}\frac{dT}{dP} = \frac{\beta_T}{\beta_T + \beta_\rho} = 0.5.
\end{equation}
Thus $\nabla_{ad} < \nabla$. 
Furthermore the Brunt-V\"ais\"al\"a frequency
in the radial direction can be calculated via (see Kippenhahn
\& Weigert 1990):
\begin{equation}
N^2(r) =  - \frac{g}{H_d}\left(\nabla_{ad} - \nabla\right).
\end{equation}
Here $g$ is the  radial component of gravity (not its absolute value as in 
Kippenhahn \& Weigert). 
At a local radius $R_0$ centrifugal force and gravity may 
cancel each other. But if a parcel of gas is radially perturbed
in a viscosity free disk, then it conserves its given
angular momentum $\Omega_0 R_0^2$. It will therefore feel an
effective gravity of
\begin{equation}
g = \Omega_0^2 \left(R_0^4 R^{-3} - R_0^3 R^{-2}\right),
\end{equation}
which can be expanded for small deviations $a = R - R_0$ into
\begin{equation}
g = - \Omega_0^2 a.
\end{equation}
Radial gravity follows locally  basically the same dependence as
the vertical component of gravity.
Thus it follows
\begin{equation}
N^2(r) = - 0.2 \Omega^2,
\end{equation}
which is obviously negative and hereby convectively unstable.
Interestingly, a  positive gradient in entropy also leads to this
situation, but in this case the flow would be unstable to perturbations
with $a<0$.

Nevertheless, the rotation prevents the disk from being 
convectively unstable in the radial direction,  which can
be seen from the baroclinic version of the Rayleigh criterion
(e.g.\ Solberg-H{\o}iland criterion)
\begin{equation}
\frac{1}{R^3}\frac{\partial R^4 \Omega^2}{\partial R} - g\frac{\partial ln P}
{\partial R}\left(\nabla - \nabla_{ad}\right) > 0
\end{equation}
which in the case of a Keplerian protoplanetary accretion disk is:
\begin{equation}
\Omega^2 - N^2 H_D \frac{1}{P}\frac{\partial P}{\partial R}>0.
\end{equation}
With $p \sim \rho T \sim R^{-2}$ it follows
\begin{equation}
\Omega^2 \left( 1 - 0.4 \frac{H_D}{R}\right) > 0,
\end{equation}
which gives a value
of about $0.96 \Omega^2$, which is a sufficient condition for stability
if only axisymmetric perturbations are permitted. If non-axisymmetric
perturbations are allowed it is only a necessary condition for stability.
Such a situation is a typical onset to the formation of non-axisymmetric
baroclinic instabilities (meanders, vortices) 
as we discussed before (see Tritton \& Davies 1985).

A detailed linear or non-linear,  possibly local,  stability 
analysis of the global baroclinic
instability in the context of accretion disks 
still has to be performed to get a sufficient criterion 
on the instability.  Such an analysis should not be confused with the
global stability analysis of a local baroclinic instability
as in Lovelace, Li, Colgate \& Nelson (1999) and Li, Finn, Lovelace, \& Colgate (2000).

\section{Changes in TRAMP}
The code {\tramp} ({\bf T}hree--dimensional 
{\bf RA}diation--hydro\-dy\-na\-mic\-al {\bf M}odelling {\bf P}roject)
was introduced and extensively discussed by Klahr et al.\ (1999).
The equations used here are the same as in that paper.
However, two additional features have been added to the code, which
are crucial for the work presented in this paper. 
\subsection{Shearing Disk}
As already mentioned in Klahr et al.\ (1999), rigid wall boundaries 
in the radial direction can affect the flow-pattern and are not the
perfect choice for simulations of disks. SB96
use the ZEUS code described in Stone \&  Norman (1992) in the
shearing sheet approximation (Hawley et al.\ 1995).
This approximation is only possible for pseudo-Cartesian coordinate
systems which themselves strongly limit the scale of the computational
domain, i.e.\ the box size has to be small in comparison to the local
radius. A direct consequence of the pseudo-Cartesian coordinates is
that there can be no global radial gradient of the density, temperature
and entropy. Thus, local studies of the global baroclinic instability are
impossible.

In order to overcome this problem, \tramp calculates on a
spherical ($R, \theta, \phi$) grid; thus large portions of the disk can be simulated, 
and one is only restricted by the computational effort required. 
We put forward a new set of radial boundary conditions that we
call {\em shearing disk} boundaries, which in fact correspond to 
a shearing sheet for spherical coordinates and disks.
In the shearing sheet approximation one uses simple periodic boundary
conditions, i.e.\ the values for the ghost-cells are determined 
from the value of cells on the opposite side of the computational
domain in a simple fashion [e.g. for the density $\rho(0)=\rho(NX-1)$, where
$NX$ is the  maximum number of zones in a given direction]. 
In the {\em shearing disk} approximation one makes two assumptions: first, 
that the mean values
of each quantity follow a power law scaling 
with the radius $R$, which is usually
the case in steady-state accretion disks for a certain range of
radii, and second,  that the fluctuations
are proportional to the mean value. Thus it follows
for the two inner and two outer ghost cells:
\begin{eqnarray}
\rho(0,J,K)&=&\rho^*(NX-1,J,K)(R(0)/R(NX-1))^{-\beta_\rho}, \\
\rho(1,J,K)&=&\rho^*(NX,J,K)(R(1)/R(NX))^{-\beta_\rho}, \\
\rho(NX+1,J,K)&=&\rho^*(2,J,K)(R(NX+1)/R(2))^{-\beta_\rho}, \\
\rho(NX+2,J,K)&=&\rho^*(3,J,K)(R(NX+2)/R(3))^{-\beta_\rho}.\label{shearbou}
\end{eqnarray}
The ($*$) indicates the implementation of the global shear in a manner similar to that given in SB96 and C96.
With the assumption that the inner and outer boundaries each move with the 
local Keplerian speed, it follows that there is a mean angular offset 
between inner and outer grid at time $t$ of 
$\delta \phi = (\Omega_{R(1)}+\Omega_{R(2)} - \Omega_{R(NX)} - \Omega_{R(NX+1)}) t/2$.
This offset translates into an integer offset $dK = ABS(\delta \phi / d\phi)$ 
defined via the angular width of a grid cell $d\phi$ 
and two interpolation factors 
$C_1 = mod(\delta \phi, d\phi)/d\phi$ and $C_2 = 1 - C_1$:
\begin{equation}
   \rho^*(K) = C_2 \rho(K+dK ) + C_1 \rho(K + dK - 1).
\end{equation}
 
The value for $\beta_\rho$ is $1$ in all 3D simulations  as we are 
performing calculations around $R=5AU$ where the surface density
is almost constant ($\beta_\Sigma = 0$). For the azimuthal frequency ($g = \dot\phi = \omega$) $\beta_\omega = 1.5$
is obvious. $\beta_{T} = 1.0$ in the 3D simulations for the temperature follows from 
the assumption of a constant relative pressure scale height
 $H_p/R={\rm const}$, 
which is necessary to make inner and outer 
vertical stratification fit in polar coordinates. 
The 2D simulations use $\beta_{T} = 0.0$ in the isentropic case (model 2)
but $\beta_{T} = 1.0$ in the baroclinic unstable case (models 3,4 and 5).
Model 6 uses open boundaries rather than shearing-disk conditions. 
Finally $\beta_g = 1.5$ for the polar velocity component ($g = \dot\theta$) assumes the 
fluctuations to be proportional to the local speed of sound.
For the radial velocity $u_r$, $\beta_u = -2 - \beta_\rho$ is 
chosen to conserve the mass flux
over the radial boundaries in a spherical coordinate system where 
$\rho u_r R^2 = {\rm const}$. It was tested that kinetic
energy flux  and momentum flux are not artificially amplified
by the boundary condition (see model 2). We damp the radial velocity 
component of the inner and outer four
grid cells by a factor of $f_d = 0.05$ each
time step ($u_r := u_r (1.0 - f_d)$) in order to prevent the model from creating artificial
resonant oscillations in the radial direction.
These radial waves can even occur in 1D radial models if there
is no damping. Model 2 (see below) is our basic test that the treatment
of our boundary conditions does not lead to a numerical instability. 
Angular momentum is not conserved, which allows the possibility of a
net transport of angular momentum into or out of the computational domain by turbulence, 
and thus of driving an accretion process or suppressing it, depending 
on the direction of the transport.

This new method was extensively tested
and will be the subject of an upcoming paper where its properties will
be compared to other choices of boundary conditions under
various physical disk situations.
 
Using this new numerical method we combine the virtues of local 
simulations (good resolution and moderately large time-steps) with
the possibility of treating global systematic gradients, previously 
exclusively reserved for global simulations.


\subsection{Reynolds Averaging}
As an addition to the earlier \tramp version there is now the possibility
to measure the transport properties of the turbulence, i.e.\ the
correlations of the fluctuations especially for the radial transport
of angular momentum. A similar method was already used in C96
as well as in SB96. This turbulent stress tensor $T$
is then scaled by the square of the sound-speed,  representing the
pressure at the midplane. The $T_{r\phi}$ component corresponds
to the angular momentum transport which corresponds to the classical
$\alpha$ value, or in other words,
an $\alpha$ is a measure of how much viscosity would be needed in order to
simulate the turbulent angular momentum transport by viscous 
diffusion of angular momentum in a laminar flow.
As we also investigate other components of the stress
tensor we define:
\begin{equation}
\alpha = A_{r\phi} = \frac{<u_\phi' (\rho u_r)'>}{\bar\rho c_s^2}
\label{alpha.ref}
\end{equation}
with
\begin{equation}
<u_\phi' (\rho u_r)'> = <\rho u_\phi u_r> - \bar u_\phi <\rho u_r>,
\end{equation}
where $< >$ or a bar represents a spatial and time average.
We use the symbol $A_{ij}$ because
the symbol $\alpha_{r\phi}$ is generally used  for
the helicity of the velocity field.

A detailed discussion of the measurements of the Reynolds stresses
including $A_{r\theta}$ and $A_{\theta\phi}$
for compressible turbulence shall be discussed in an upcoming paper
(Klahr \& R\"udiger in preparation).
In a similar manner we measure the strength and isotropy of the 
turbulence via the determination of the on-diagonal terms of
the stress tensor $T_{rr}$, $T_{\theta\theta}$ and $T_{\phi\phi}$ which, scaled
by the square of the sound speed,  are three different indicators of the 
square of the turbulent Mach number: 
\begin{equation}
A_{rr} = \frac{<{u_r'}^{2}>}{{c_s}^2},
\label{alpha_rr.ref}
\end{equation}
\begin{equation}
A_{\theta\theta} = \frac{<{u_\theta'}^{2}>}{{c_s}^2}
\label{alpha_tt.ref}
\end{equation}
and 
\begin{equation}
A_{\phi\phi} = \frac{<{u_\phi'}^{2}>}{c_s^2}.
\label{alpha_pp.ref}
\end{equation}
In the same way we determine the mean density fluctuation in  the
flow by $\frac{<{\rho'}^{2}>}{\bar\rho^2}.$
We define $M = \sqrt{A_{rr}+A_{\theta\theta}+A_{\phi\phi}}$ to be
the Mach number in our models. 

\section{3D Radiation-Hydrodynamical Simulations}
The inviscid 3D simulations of thermal convection in disks 
(Klahr \& Bodenheimer 2000)
showed radially outward directed angular 
momentum transport if the section of the 
disk was large enough. In the following we
show that the angular momentum transport is not 
dominated by the vertical
thermal convection but by the hydrodynamical
turbulence in the $R-\phi$ plane of the disk
which results from the radial entropy gradient
self-consistently introduced by solving the
energy equation in our radiation hydrodynamical
code.

As a continuation of our previous work (Klahr et al.\ 1999),  we have 
performed  simulations of 3D chunks of protoplanetary
accretion disks. We have tested  the influence of viscosity,
artificial heating and boundary conditions extensively.
As highly viscous flows tend to become axisymmetric
(C96; Klahr et al.\ 1999), we were especially interested
in simulations with high Reynolds numbers as in SB96.
In model 1 we switch off the application of the viscous 
forces in the momentum equation but keep the heating, corresponding
to  $\alpha = 0.01$,  from the local 
viscous dissipation function ($\Phi$) in the energy equation
\begin{equation}
\Phi = \left(\nabla{{\bf T}}\right){\vec{u}},
\label{Eq_Dissi}
\end{equation}
where ${\bf T}$ is the viscous stress tensor.
For the viscosity $\nu$ we adopt $\nu = \alpha c_s^2 / \Omega_{Kepler}$.
For more details we refer to Klahr et al.\ (1999). 
In an additional simulation (model 1b) we
forgo the artificial heating and show that a self-consistent heating
of the disk via the dissipation of compression waves,  especially shocks,
and the release of gravitational energy 
can lead to the same baroclinic effects as the artificial heating in model 1.
Artificial heating simplifies the numerical effort, as it stabilises the
vertical and radial structure of the disk and thus allows for longer integration times for
the mean values (see Table 1).

Artificial von Neumann \& Richtmyer (1950) viscosity (see also Stone \& Norman 1992) for the proper treatment shocks is included, as
is flux-limited radiative transport with a simple dust opacity  ($\kappa =   2 \times
10^{-4} T^2$ cm$^2$ g$^{-1}$). 
The goal was to investigate the influence of our more global 
approach, especially a wider azimuthal range, on the simulation results.

The particular models we present here 
(model 1 \& 1B; see Table  1) cover a radial range from
$3.5$  AU to $6.5$ AU, a vertical opening angle of $\pm 7\arcdeg$, and 
an azimuthal extent of $90\arcdeg$. They use only 20 grid cells in
the vertical direction but 51 in radius and 60 in azimuth.
A follow-up paper will deal entirely with 3D models at various
resolutions. At the moment it is sufficient to say that the results
of model 1 are typical and remain the same at much higher resolution.
The benefit of model 1 is that it  can easily
be evolved for 68 orbits at the outer edge of the simulation
within sixteen CPU hours on a Cray T90. In future papers
we will discuss the influence of artificial heating in contrast
to self-consistent heating due to shock dissipation in greater detail. 
Calculations as model 1B indicate at least that numerical dissipation 
might be a problem as it removes kinetic energy from
the system without
heating the gas. Thus there is less heating
to maintain the initial vertical and radial temperature structure of
the disk.
Higher resolution models will have to investigate the
effect of numerical dissipation. 

   Figure \ref{fig1.ref} shows the three-dimensional rendering of the
vertical mass flux distribution after model 1 approached 
a quasi-steady state. The azimuthally extended convection cells
are slightly twisted and not completely axisymmetric.
In Figure \ref{fig2.ref} the normalised Reynolds 
stresses and corresponding Mach numbers of the turbulence 
for model 1 are given, time-averaged
over 68 outer orbits.
The radial mean value for the angular momentum transport
corresponds to $\alpha = A_{r\phi} = 2 \times 10^{-3}$, 
a quite high value, and the tendency towards positive values is obvious.
The radially  varying Reynolds stresses reflect their
connection to locally and temporarily confined flow structures
and events. Only a much longer time for averaging,  as is done 
in the 2D simulations, can produce a more homogeneous distribution.
Based on the measured stresses one can estimate a viscous time-scale
of $\approx \alpha^{-1} = 500$ orbits, while for technical
reasons only 68 are possible in one run.
The on-diagonal terms of $A$, which characterise the strength and
isotropy of the turbulence, show that the fluctuations in radial and 
azimuthal velocity are
almost sonic at some radii,  and that the fluctuations in 
the vertical (polar)  velocities are more than an order of magnitude
smaller.  This difference reflects the fact that (1) the radial and azimuthal
velocity fluctuations are not driven by the vertical thermal convection, and (2)
there is no isotropic (3-D) turbulence present.
   The resulting
Mach number of the turbulence, the combination of the
strengths  in three dimensions,  is close to one. The relative density
fluctuations are in the mean as strong as $10\%$.

In contrast to this result, with positive values of the Reynolds
stress, our corresponding  axisymmetric 
2D simulations, with the same physical assumptions and parameters as in 
model 1, always showed clearly negative 
Reynolds stresses.
The dissipation-free axisymmetric Euler equation would
predict precisely zero average stresses,  which is a consequence
of strict angular momentum conservation, but radiation hydro
calculations are by nature not dissipation free even if one has
no viscous forces. Thus the negative $\alpha$ values are
a result of the heating and cooling of the gas.

Looking at the flow-pattern and density distribution in the
$R-\phi$-plane (Fig.\ \ref{fig3.ref}), 
we do not see the convection cells from Figure 
\ref{fig1.ref}. 
What we see are vortices, vorticity waves (e.g.\ Rossby waves, baroclinic waves), 
and pressure waves propagating in the flow. 
These are the sources of the positive
Reynolds stress. They are counteracting the negative Reynolds stresses
generated by the thermal convection. In the turbulence terminology
one would possibly identify geostrophic turbulence (irregular waves; 
see e.g.\ Tritton \& Davies 1985). This can be seen in
the spectral density distribution of the flow in the azimuthal
direction (Fig.\ \ref{fig4.ref}). Only at the
smallest resolved scales ({\em i.e.\ meso scale}) the slope approaches 
the Kolmogorov spectrum ($k^{-5/3}$).
At larger scales the slope is even steeper than for earth atmospheric 
geostrophic turbulence ($k^{-3}$) which indicates how strong small-scale
turbulence is being pumped into  the smallest possible wave numbers.
In this calculation the minimal $k$ is four as we calculate only
a quarter of the disk. However 360$\arcdeg$ full circle simulations (see model 6) 
 indicate that
$k=1$ is the mode where the energy will pile up. The dissipation in
this calculation thus does not only occur at the smallest scales
as in 3D incompressible turbulence
but at all scales, especially at the large scales where shocks form.
The idea to drive the accretion process via shocks was already
suggested by Spruit (1987), but in his barotropic models the
shocks could not form without the presence of an external perturber.

The Mach number of the flow is the smallest in the midplane 
(compare  $M_{max} = 0.6$ in Fig.\ \ref{fig3.ref}  with 
$M=0.5$ in Fig.\ \ref{fig2.ref}). Velocities rise,  
eventually beyond the sound speed, with height above the midplane.

From this result several questions arise.
(A): Why did neither Stone \& Balbus (1996) 
nor Balbus, Hawley \& Stone (1996)
observe these positive Reynolds stresses
and violent turbulence?
(B): What is the source for the vorticity and
turbulence in the $R-\phi$ direction?
(C): Especially, how can we 
be sure that we are not observing  boundary effects?
We will explicitly answer these questions in \S 7.

The strategy in order to identify the source for
turbulence and angular momentum transport was to
simplify our model step by step by removing
physics from the simulation and to wait for the turbulence 
and Reynolds stresses to disappear. The first effect we
found was that whenever we decreased the azimuthal extent
of the computational domain below a critical value of
$\approx 15$ degrees, the Reynolds stresses switched to negative values, as 
in 2D axisymmetric calculations (Klahr \& Bodenheimer 2000b). 
As the SB96 calculation only covers roughly 1 degree,  this was a first
hint that our findings might result from the more global
simulation.

\subsection{No artificial heating}
   The artificial heating in model 1  has the benefit that
there is a well defined laminar equilibrium state for
the disk. This is ideal for any stability 
investigation as well as for the setup of a numerical model.

   But one can argue that this artificial heating is 
inconsistent with the lack of dissipation of kinetic energy.
Consequently, the findings with model 1 could result purely 
from the artificial heating, or,  even worse,  from the specific
shape of the dissipation function (see Eq.\ \ref{Eq_Dissi}).
Thus, we repeated the simulation of model 1, but switched
off the heating completely. This simulation is numerically more complicated 
and more expensive than model 1 as the initial state is 
far from equilibrium and the disk undergoes strong fluctuations
in density and temperature. Only 14 orbits could be performed during a
10-hour run on the Cray T90. 
The disk vertically shrinks by 20 percent
before it reaches a new self-consistent state with a 
midplane temperature of 40 K, down from  150 K  in the initial state.

If the temperature changes by a factor of almost 4 (150 K $\rightarrow$ 40K) 
one could expect the disk height ($\sim$ pressure scale height ) 
to change by about a factor of 2.  
But the factor of 2 in scale height is only true for vertically isothermal disks.
Our disks have a vertical temperature profile that becomes flatter as the temperatures
decrease because  the opacities also  decrease (proportional to $\kappa
\propto T^2$). This means that the high-temperature disk has
temperature, pressure and density profiles that fall off much more steeply  than the pressure
scale height estimated for the midplane temperature might suggest.
The cooler disk on the other hand is closer to an isothermal disk and
thus does not shrink in proportion to the root of the midplane temperature.

The vertical grid structure was readjusted during this cooling:
the initial vertical opening angle was $\pm 7\arcdeg$, but 
during the vertical shrinking the gradients above the disk became
too  steep for the numerical technique to work. Thus we constructed a new vertical 
grid  $\pm 5.5\arcdeg$ also with  20 grid points and interpolated the
values linearly from the old onto the new grid.

The newly gained state of 40 K was then maintained for
more than 100 orbits (see Fig.\ \ref{fig5.ref}, which is  similar to 
Fig.\ \ref{fig3.ref}). 
The disk turbulence
was transonic (see Fig.\ \ref{fig6.ref}) and the heating 
was highly non-homogeneous in space and time, in contrast to 
model 1 where it was assumed to be smoothly distributed.
 But nevertheless the general result was the same:
the disk develops a radial entropy gradient, turbulence and
vorticity. In addition, the accretion of matter which results
from the positive Reynolds stresses feeds energy into the system
which is radiated away after the turbulence is dissipated in shocks.

   These self-consistent simulations will have to be continued in 
our future work. The simulations are numerically much more
unstable than the heated or polytropic models. But in the end
only they can provide realistic predictions for density, temperature, 
turbulence,  and the evolution of accretion disks.

\subsection{Radial entropy gradient}
   Another glance at Figure \ref{fig3.ref} 
shows that temperature (contour-lines)
and density (colors) are not perfectly aligned. Thus also pressure
and density gradients do not point in the same direction, a clear
indication of a baroclinic flow, where the entropy decreases with
radius. To test the relevance of
this possible instability for the generation of the observed turbulence 
we removed the vertical structure of the disk and got rid
of the radiation transport. 
Thus the 3D density $\rho$ was replaced by the surface density $\Sigma$
and the 3D pressure $p$ by the vertically integrated pressure $P = \Sigma T$
(in dimensionless units).
In Figure \ref {fig7.ref} 
we plot the $K \sim T \rho ^ {1-\gamma}$ values in the midplane (before
model 1 became turbulent) as $K/K_{max}$. If one measures
this $K$ value in the turbulent state of the disk,
the fluctuations of $K$ are too strong, which hides the general
trend. The next section describes the results 
of using such an entropy distribution in a two-dimensional ($R,\phi$) disk. 

\section{2D Simulations}
The flow characteristics in the 3D simulations, e.g.\ the turbulence
cascade and the small vertical velocity fluctuation, indicated
that the disk instability must be a 2D effect. In order to verify
this thesis we removed the vertical structure and performed
2D $(R-\phi)$ simulations.

Both of the following models were first developed as 1D radial
axisymmetric models with density slope $\rho \sim R^{-1}$
($\rho[1AU] = 10^{-10}$ g cm$^{-3}$), 
which corresponds, in the 2-D flat disk, to $\Sigma= $ const ($\approx 300$ g cm$^{-2}$). 
The initial temperature was then chosen to be either constant 
              ($T = T_0$) as a function of radius  or to follow a power law ($T = T_0
              \left (R/R_0\right)^{-1}$). The value for $T_0$ was
              adjusted to create a local pressure scale height of $H/R = 0.055$
              at the mid-radius $R_0=5$AU, to match that in the 3D radiation 
              hydro calculations. For obvious reasons this pressure scale height 
              condition is only fulfilled at $R_0$ for the constant temperature
              case but for all radii in the $T \propto R^{-1}$ case.
In the constant temperature  model 
$H/R$ varies as $R^{0.215}$, which
means that the relative pressure scale height increases slightly
with radius.  Both models use 
an identical computational domain with radii between 
4.0 and  6.0 AU and an azimuthal extent of 
$30\arcdeg$, with a moderate resolution at $64^2$ grid cells. The boundary
conditions are identical to those of model 1: a shearing disk plus  damping of the
radial velocities in 4 grid cells 
near the inner and outer boundaries. 
The artificial viscosity is the same as in model 1.    
Both models are initially kicked
by a random density perturbation with 1$\%$ amplitude: $\rho := \rho * (0.99 + 0.02 * RAND)$, where $RAND$ is
a uniformly distributed random number between 0 and 1.
Afterwards the system was allowed to evolve
freely.
   We advect the internal energy and calculate the $PdV$ work arising from
compression and shock dissipation in the same manner as in the 3D calculations.
For the time being there is no radiative cooling, because cooling would
immediately change the temperature profile and we are interested in the
behavior of given temperature distributions. Models with radiative
cooling shall be dealt with in future work.

\subsection{Model 2: radially constant entropy: ($T = const$)}
This model developed no turbulence. The initially induced turbulence 
decayed rapidly with an e-folding time for the kinetic energy in $R$ of 
$\approx 4.3$ orbits (see Figs.\ \ref{fig8.ref} and
 \ref{fig9.ref}).
Nevertheless, the initially introduced turbulence 
always transported angular momentum outward, never inward (see Model 2B).
Figure \ \ref{fig8.ref} shows the surface density distribution and flow field after about 90 orbits.
The surface density is almost constant ($299.955<\Sigma <300.125 $ g cm$^{-2}$). The arrows indicate that there
is still small noise in the velocity field with a Mach number of about  $10^{-4}$. In Figure 
\ \ref{fig9.ref} the evolution of the
total kinetic energy in both coordinates, and the enstrophy of the flow are plotted, in units
of the initial values. 
The overall mean mass accretion rate is given in solar masses per year. 
Here and in the following models, a negative value of \.M indicates
accretion towards the star. 
Enstrophy is the integral square
of the local vorticity. One clearly sees that the kinetic energy  as well as
enstrophy are 
decaying  as is expected for
a barotropic disk. Kinetic energy in $\phi$ and enstrophy are calculated 
from the deviations from the Keplerian profile.
The residual finite values of these quantities
result from the laminar (not strictly Keplerian) steady state flow.
The resulting mean turbulence, measured by the strength 
of the components of the stress tensor (see Fig.\ {\ref{fig10.ref}}) is also very low.

We also tested the effect of stronger initial perturbations.
For model 2B we gave model 2 an initial random density fluctuation 
of $\pm  50 \%$.
Still the turbulence rapidly decayed as expected. But we have to
stress that the turbulence in this barotropic simulation
had already the property of effectively transporting angular momentum outward.
The Reynolds stresses during the first orbit (see Fig.\ {\ref{fig11.ref}}) are positive ($<\alpha = A_{r\phi}> \approx  3.0 \times 10^{-4}$), nevertheless rapidly decreasing afterwards.

We conclude that positive Reynolds stresses
are a common feature of hydrodynamical turbulence in disks,
and not exclusively a result of baroclinic simulations.
This is also a result of Hawley et al.\ (1999),  which shows 
that positive Reynolds stresses are correlated to the decay
of turbulence.
Only thermal convection, which must not be confused with
isotropic turbulence, has a tendency of transporting angular
momentum inward. Nevertheless, in barotropic disks hydrodynamical
turbulence can not sustain itself and decays.

\subsection{Model 3:  radially  varying entropy: ($T \sim R^{-1}$)}
Using a different initial state leads to a completely different result.
The flow becomes turbulent within a few orbits, with kinetic energy rising
exponentially until saturation due to shock dissipation occurs.
Model 3 (Fig.\ \ref{fig12.ref}) shows some similarities to 
model 1 (Fig.\ \ref{fig3.ref}) even though we are
comparing a 3D and a 2D simulation. After 100 orbits the 
surface density shows deviations from
axisymmetry ($286<\Sigma <318$  g cm$^{-2}$).
The velocities are $10^2$ times stronger than
the ones in model 2. 

The initial instability grows quickly  (Fig.\ \ref{fig13.ref}). The kinetic energy
in the radial direction grows by a factor of $10^3$ within 40 orbits.
This translates into a characteristic growth time of $\approx 5.0$ orbits.
The other components, especially the enstrophy,  grow more slowly 
but
do not saturate as quickly 
 as the radial kinetic energy (which is also slightly
damped for numerical reasons [see \S 3.1]). 
The strong rise of enstrophy ($\approx$ vorticity generation) is an
indication that a baroclinic instability is at work.

The angular momentum transport and also the on-diagonal components of the
 stress tensor (Fig.\ \ref{fig14.ref})
are orders of magnitude stronger than in model 2.
 They are weaker than in models 1 and 1B, 
which may result from the smaller computational domain, 
especially in the azimuthal direction, artificially limiting
the wave numbers of the instability.
The angular momentum transport ($\alpha = A_{r\phi}$) is in the mean as strong as 
$1.5 \times 10^{-4}$. This value should be 
contrasted  to the strength of the turbulence itself. An
 $A_{rr}$ and $A_{\phi\phi}$
of $\approx 10^{-3}$ correspond to a turbulent Mach number as strong as 0.05.
In isotropic turbulence this value could be used for a mixing
length model and would
then predict an alpha-viscosity of $\alpha = 2.5 \times 10^{-3} - 5\times 10^{-2}$,
 depending on the estimates
for the ``typical eddy size''. The lower value of $\alpha$  
that we obtain indicates that even when strong turbulence  develops
it does not automatically generate strong Reynolds stresses, which again
is evidence
for non-isotropic turbulence. The energy of this turbulence is drawn from the entropy
background which can maintain its profile due to the accretion of mass. 
The measured mass accretion rate of $\dot M \approx 
- 2.0 \times 10^{-9}~\mbox{M}_{\sun}~\mbox{yr}^{-1}$ is in rough agreement with the 
expectation from a viscous accretion disk model with a surface density
of $\approx 300$ g cm$^{-2}$ and an $\alpha$ parameter of $1.5 \times10^{-4}$.

The turbulence saturates due to dissipation from shocks and stays at 
a high, almost sonic, level. This level is on the long term quite
variable. 

Our simulation demonstrates  that non-magnetic turbulence can drive outward angular 
momentum transport
and is maintained itself by the resulting accretion process.
Such a conclusion has been in doubt for a long time,  as no 
instability mechanism has been convincingly demonstrated 
to work in Rayleigh-stable and non-ionized disks.

   We checked our findings on model 3 by redoing the simulation at 
twice the resolution (model 4: $128 \times 128$
grid cells) and obtained general agreement (see Fig. \ref{fig15.ref})
with a mean $\alpha = A_{r\phi}$  slightly larger than that of model 3. 
A detailed study of the influence of resolution on our results
is a current subject of our investigations and will be
part of a future paper.

   For model 5 we used also a resolution of $128 \times 128$
grid cells, but this time we used a computational domain which
was essentially twice as wide in both directions. It spans radii from
3 AU to 7 AU and  $60\arcdeg$ in the $\phi-$direction. The numerical
resolution is thus about the same as in model 3. 
In Fig.\ \ref{fig16.ref} we plot the Reynolds stresses,
which are stronger than in the previous models with $\alpha = A_{r\phi} \approx
 6 \times 10^{-4}$.
This larger scale simulation showed also the formation of a vortex (see Fig. \ref{fig17.ref}).
It is an anti-cyclonic prograde vortex with higher density and pressure than the back ground.
The vortex emits spiral waves in the ambient accretion disk.


As model 5 is so far the best compromise between resolution and
angular width, it is well suited for the examination of the  spectral density
distribution  (see Fig.\ \ref{fig18.ref}) and 
comparison with model 1 (see Fig.\ \ref{fig4.ref}).
Here one clearly sees the transition between 2D and ``3D" characteristics (even though
the calculations are actually 2-D)
of the spectrum at a wave number of about $60$, a value which was
so far observed
in all simulations. This transitional wavenumber is known as ``Rhines
blocking wavenumber''; roughly speaking, 
 for $m = k > 60$  energy only cascades down to smaller 
scales while for $k < 60$ it inversely cascades
up to larger scales. The wave number $k=60$ corresponds to about 6 degrees in azimuth or
a tenth of the radius, which is almost two pressure scale heights.
Structures that are smaller than this are not influenced by the rotation and 
apparently behave like 3D turbulence ($k^{-5/3}$), i.e.\ energy 
cascades down to smaller scales to be ultimately dissipated.
Structures bigger than this feel the Coriolis force and thus the rotation
and shear. Consequently they are forced to have 2D characteristics and an inverse
energy cascade with $k^{-3}$. The same behavior can be observed
in the earth's  atmosphere,
where small structures up to 500 km in extent (i.e. local winds) 
behave according $k^{-5/3}$, while larger structures 
(like hurricanes) follow a $k^{-3}$ distribution.

The studies on the influence of the computational domain
will have to be continued in future investigations.
The same is true for parameter studies on the initial
distribution of density (as well as $\gamma$ and temperature) 
in the disk models and the influence on $\alpha$.
The fact that we found a working hydrodynamical instability for disks which generates
Reynolds stresses with the right sign and reasonable $\alpha$-values,
seems to support the paradigm of an viscously driven accretion process. Considering 
the limitations of our numerical method, we could argue that a better method or higher
resolution could also lead to even stronger turbulent viscosity, as only the
sound speed seems to be a natural upper limit for the velocity fluctuations.

Other flow characteristics besides the Reynolds stress
are very strongly developed. First, the 2D-turbulence itself  produces a
significant turbulent (r.m.s.) pressure which is not accounted for 
in $\alpha-$models. Second, the vorticity
and the deviation from the mean Keplerian profile are significant 
features, which also
cannot be handled in diffusion models. And finally even though $\alpha-$
models might assume that
there is some underlying non-axisymmetric turbulence, 
they never would predict the
formation of large scale flow features like long-lived and growing vortices of 
several scale heights diameter. Even if such features were assumed 
to be present initially, $\alpha-$ models would smear
them out and not amplify them 
(Godon \& Livio 1999b).

We have now demonstrated the capability of the global baroclinic instability
to form non-axisymmetric structures and vortices in the local simulations
with periodic boundaries. The next question to be investigated concerns
what would happen if 
the global shear were not fixed by the radial boundary conditions but were
free to evolve.  Thus
we set up a global baroclinic simulation to see how a real disk would 
behave and to what extent it could still be described by an $\alpha-$model.

\section{A global 2D simulation}
Our global simulation shows that all the findings from the
local models (model 3,4 and 5) can be confirmed. Actually the
Reynolds-stresses are even stronger than in the local simulations,
as smaller wave numbers in the azimuthal direction can be resolved.
The wave number $ m = 1 $ seems to be the preferential mode of
the instability. A further discovery was the self-consistent 
formation of long-lived anti-cyclonic vortices as a direct result
of the global baroclinic instability. This may have major relevance
for the formation process of planets.

The simulation (model 6) covers the entire $360\arcdeg$ of the circumference of the
disk and a radial section between 1 AU and 10 AU. The grid measures
$128\times128$ grid-cells, which are radially logarithmically distributed. 
We can thus resolve azimuthal wave numbers between 1 and 64. 
The boundary conditions in the radial direction were
changed from periodic to simple non-reflecting outflow conditions (vanishing 
gradients),  not allowing for inflow. 
 As a result of this change we also could drop the damping of the radial
   component of the velocity close to radial boundaries.
The density distribution 
again was $\rho \propto  R^{-1}$ (constant $\Sigma \approx 300$  g cm$^{-2}$), 
 and the temperature distribution was $T \propto  R^{-1}$, thus we have a
 baroclinically unstable situation  as in model 3, which results from $H/R = 0.055$.
The model was first run into a stable 1D axisymmetric
state, where the residual velocities were less than 
$10^{-4}$ cm s$^{-1}$. 
Without a symmetry-breaking instability and turbulence generation, this disk 
can not evolve and would stay perfectly laminar forever, as in 
 the ``dead zone'' described by  Gammie (1996).

The initial density distribution was then 
perturbed by  random noise of amplitude only 0.1$\%$. 
The initial state is
practically axisymmetric.
Figure \ \ref{fig19.ref} illustrates the evolution of the flow in two
space dimensions, over the full 360$\arcdeg$.  
After the first 1 orbit (30 yr at 10 AU; Fig.\ \ref{fig19.ref}a) only little
structure has evolved.  But with time a prominent anticyclonic vortex forms,
which reflects the assumption that $m=1$ is the preferred mode.
Intermittently a  second vortex also forms,  and we assume that
their number is just limited by the narrowness of our disk and 
a lack of matter.
The vortex grows in mass and
propagates radially outward,
 possibly as a result of the gradient of background
vorticity and the fact that anticyclonic vortices 
are a local sink for angular momentum in the global vorticity field.
Anyway, as there was never a radial drift of
vortices in barotropic flows reported, this effect might
also be linked to the baroclinic features of the flow.
A detailed investigation of this effect still has to 
be performed.

Figure \ \ref{fig20.ref} shows the Reynolds stress and 
turbulent Mach number averaged over the orbits from 430 to 500. 
We see that the Reynolds stress does not
disappear as we remove the ``shearing disk'' boundaries. 
The stresses are comparable in the main part of the disk to those in 
models 3, 4, and 5  and only deviate at the physical edge of the disk where 
density and sound speed drop by orders of magnitude. 
We can conclude that the ``shearing disk'' 
boundary condition does not significantly affect the results. 

Figure \ \ref{fig21.ref} shows the situation after about $ 10^4$ years
in real Cartesian coordinates to give an impression of the global
nature of the simulation.
Figure \ \ref{fig22.ref}  shows the flow pattern in more detail 
in ($R,\varphi$) coordinates at the end of the simulation. 
A huge vortex has formed which has a factor 4 over-density with respect to the
ambient disk and a factor 2 over-density with respect to the initial local
surface density. It is a high-pressure anticyclone that has the property of
collecting solid material in its center (Tanga et al.\ 1996; Godon \& Livio 1999b).
At the same time the over-dense blob inherits a substantial fraction of the 
disk gas, which is not confined  by self gravity but only by the pressure
gradient generated by the anti-cyclonic (less pro-grade) rotation. 
While the initial nebula (from $1-10$ AU) had a mass of about 
$10^{-2} M_\sun$, there are only $8 \times 10^{-3} M_\sun$
left after $10^4$ years which corresponds to a mass loss (radially inward and
outward) of $2.0 \times 10^{-7}~ \mbox{M}_{\sun}~\mbox{yr}^{-1}$.
We cannot tell in this simulation how fast mass is being accreted onto the star
 as our computational domain ends at 1 AU. 
The red blob collects a total of $\approx 10^{30} $  g (170 M$_\earth$ 
or 0.5 Jupiter masses [M$_{\rm J}$]).
 Without further addition of mass or cooling 
this object is not gravitationally unstable, as the Toomre $Q$ in the
disk is about 10 and the local Jeans mass in the condensation is 
about 2.5 M$_{\rm J}$. 

We see that even in a disk which 
is not massive enough to fragment into planets or brown dwarfs (Boss 1998),
a kind of pre-protoplanet can form simply as a result of baroclinicity and
the resulting vorticity. The object, which is not yet gravitationally
bound, could evolve into a planet in one of two ways: 
(1) it could  efficiently collect solid particles in the center and
wait until the critical mass for gas accretion is reached, or (2) it could 
concentrate enough gas and cool down efficiently to become
gravitationally unstable. In either scenario the time scale for planet 
formation  will
be shorter than in cases without the vortices in the disk.
Additionally the vortex scenario can explain why there could be a solid
core in a
planet even it formed by gravitational instability rather than by accretion
of solid material to form a core followed by gas capture. 

\section{Conclusions}
The global baroclinic instability is found to generate
turbulence in disks and drive an accretion process.

In general we found numerically that
isotropic turbulence in the $R-\phi$ plane of the disk has
the property of transporting angular momentum radially outward.
But only the global baroclinic instability seems to provide
a reliable source for this turbulence in the first place.
Thermal convection in the vertical direction of the disk
is not necessary for this effect.

These results seem to be unaffected by the type of simulation.
Whether 2D or 3D, whether polytropic, artificially heated or
not heated, whether open boundaries or shearing disk, all
simulations lead to the same conclusions.
Nevertheless, only global 3D non-heated models will have
the credibility to predict exact $\alpha$-values.

We showed 
that a protoplanetary disk  with a
density and temperature distribution which cannot be described
by a single polytropic $K$ is not barotropic and is possibly 
unstable against non-axisymmetric
perturbations. 
This non-isentropic  situation (with a radial entropy gradient) is ultimately
a consequence of the radially decreasing gravitational force.      
The vertical component of gravity pushes the disk together and determines
the pressure. Thus it is natural for non-isothermal  disks to
be baroclinic.

We demonstrated numerically in a simple 2D simulation that 
a baroclinic disk is unstable and develops strong geostrophic
turbulence while a barotropic disk is perfectly stable.

Now we can answer the questions raised in \S 4.
(A): Balbus et al. (1996; see also Hawley et al.\ 1999) could not observe 
this kind of instability
as their simulations were barotropic. The simulations in SB96 allowed
for local baroclinic effects but no global entropy gradient was present.
They used the shearing-box 
approximation; thus $\beta_\rho, ~\beta_T = 0$ and $\beta_K$ is 
automatically zero. 
(B): The instability is purely hydrodynamical with an initial e-folding 
time for
the growth of about
5 orbital periods for an entropy gradient with $\beta_K = 0.57$. It occurs naturally
in rotational shear flows
when surfaces of constant density are inclined versus surfaces
of constant pressure. A detailed stability analysis does not
exist yet, but this is true for a lot of turbulent
situations. Indications are, however, that the lowest-order
modes have the fastest growth rates.  It is also not known yet whether we
are  observing a linear or a non-linear instability operating in the disks. 
(C): The barotropic  shearing-disk simulations as well as the baroclinic 
open-boundary simulations  indicate that
the ``shearing disk'' boundary conditions are not responsible
for the turbulence. 

   Thermal convection is only indirectly related to the baroclinic
instability,  as convective and radiative transport are responsible 
for the radial temperature distribution and therefore the 
 radial entropy gradient. 
An isothermal disk would always be barotropic,  but
we are considering only the situation 
where the optical depth is larger than one, so the disks are not
isothermal. 
The optical depth greater than one is also necessary
for thermal convection, but a vertically
convectively stable purely radiative disk 
can also establish a radial entropy gradient.
If convection is present,  then it can be important in generating 
the initial non-axisymmetric perturbations induced in the Keplerian
disk which are necessary to set off the baroclinic instability. 
 On the other hand convection produces negative Reynolds
stresses;  however these turn out to be 
 orders of magnitude weaker than the ones created by
the baroclinic instability.

The simulations also generate vorticity, as can be made plausible by 
a simple argument. Equation (\ref{eq:unstable}) shows how a density fluctuation in
the $\varphi$ direction leads to a pressure gradient that does not
exactly line up with the density gradient. Thus 
imagine two parcels of gas on the same orbit around
the central object. When they get pushed apart in the azimuthal direction,
then the pressure will try to restore the previous state by pushing them
together again. In a barotropic disk they  perform a damped oscillation,
 and the perturbation decays. 
But in a baroclinic disk, assuming that the perturbation occurred
only along the azimuthal direction, the gas pressure will now not only push
the parcels azimuthally together again, but also will drive the gas parcels
slightly radially outward. For two reasons they are then driven 
 back to the lower radius.
First, there is  the gas pressure at the larger radius,  and,  second, 
 they do not have enough angular momentum to stay on that higher orbit.
When they  return to their lower orbit, they will fall behind their
original azimuthal position, as the gas on a lower orbit has a larger
angular velocity then the one on the outer orbit. And thus it is clear that 
a non-axisymmetric radial velocity distribution is created ---
$\partial v_r/\partial \phi \neq 0$ --- and thus vorticity generated.
This leads
first to a ``meandering'' flow which eventually
becomes completely chaotic,  characterized by irregular waves.
Furthermore
a radially local perturbation spreads quickly in the radial direction,
as it always affects and perturbs neighboring  radii.
 Other works (e.g. Adams \& Watkins 1995) 
 have considered the behavior of vortices and their interaction 
with viscosity before, but none of them  have created them. We show here  for the first
time that they form necessarily in a realistic disk in the absence
of an underlying viscosity.

Finally,  the baroclinic instability generates turbulence with velocities close to the sound speed.
Dissipation does not occur (homogeneously in time and space) on the small 
scales as in $\alpha$ models but occurs in       
large-scale shock structures. 
The calculations strongly suggest that this baroclinic instability is a feasible
way to maintain turbulence  and outward angular momentum 
transport in protoplanetary disks,  even if the physical
conditions do not allow for MHD  turbulence.
It would follow that there is no such thing as a ``dead zone'' in protoplanetary disks.
Evidently then also the paradigm of layered accretion has to be revised. 
One also has
to take into account a 
transition zone in  a disk, which separates regions where
MHD turbulence or hydro-turbulence dominate, and in which the two
processes may coexist.  
It also would be fruitful and interesting to study 
the transition zone and in general the interaction between hydro
turbulence and self gravitational instabilities, as such a process
could be crucial for planet formation. These combined effects would be important
at the earliest stages of evolution of a protoplanetary disk, when
the disk is still relatively massive and 
material is still accreting onto the disk from the surrounding molecular
cloud. Once the disk becomes optically thick and radiative 
transfer effects become important, one can expect baroclinic effects to
occur. 

 Gravitational and baroclinic instabilities seem to have certain 
properties in common. Each of them leads to non-axisymmetric modes
and an angular momentum transport that is not necessarily describable 
by an $\alpha-$ formalism.  In baroclinic disks we are able to
measure quite reasonable, and even more significant, all positive values
for the Reynolds stresses, but it could  be dangerous to use these values
in the viscous description of an accretion disk. Our global model
shows dramatically how far a real disk can depart from the 
idealized axisymmetric laminar disk that 
evolves in quasi-steady state on viscous time-scales.
Thus, results from $\alpha-$disk models can only reflect the 
longterm average properties of disks.

 The trans-sonic  turbulence would be expected to be  
very influential on the passive dust contaminant in generating
collisions and mixing as well as concentrating the dust grains.
Nevertheless the  Mach number associated with mixing of angular momentum is almost two
orders of magnitude less than that of the turbulence, which reflects the non-isotropic
2D character of the turbulence. 
The spectral density (see Fig.\ \ref{fig4.ref} and Fig.\ \ref{fig18.ref}) indicates that there is also turbulence on the
small scales, but only resolution studies can show how much energy
really decays towards the Kolmogorov scale.
In short, a baroclinic disk is  much more 
turbulent than `viscous'.

   Our simple global 2D model already shows azimuthal density fluctuations
by a factor of four. But most up-to-date models which try to interpret observational
data use axisymmetric disks. Our global simulation could be a first step
in showing how the spectral 
(radiation) energy distribution (SED) of a baroclinically unstable disk
would look like in contrast to an axisymmetric one.

The anti-cyclonic rotating gas parcels are vortices, that
could be the precursors of planetary formation. 
They can be thought of as pre-protoplanets.  
In this connection, 
more realistic 3D models with radiation transport
could allow for higher mass concentration, as the cooling can locally decrease
the pressure. The planets could form either by concentration of dust in the centers of
the vortices, as was suggested by Tanga et al. (1996) and Godon \& Livio (1999b),
or by sufficient gas accretion onto a vortex so that it undergoes 
gravitational collapse (Adams \& Watkins 1995). This second process would
happen in a similar fashion to the model by Boss (1998) with the big difference
that the vorticity takes care of the local mass enhancement even after the disk
or even parts of the disk have become gravitationally stable.  It furthermore is an easy way
to explain a solid core for Jupiter and Saturn, as the pre-planetary embryo, the 
rotating slightly over-dense vortex, will already have accumulated planetesimals.

 In a final speculation  we suggest that there is  a connection
between UXor events (Natta  et al. 1999) and the baroclinic instabilities.
It would be logical to expect over-dense vortices to form
in disks around Ae/Be-stars. In a nearly edge-on disk, a transit of such a cloud, which has a 
 higher scale height than the material in the surrounding disk,
could obscure the stellar
light. In a viscous $\alpha-$disk these over-dense regions would smear
out in the azimuthal direction on a dynamical time-scale, which is an orbital period, but 
in a viscosity free disk, they do not only persist, they can form.

Further 
2D and 3D simulations are necessary to determine 
the proper role for the global baroclinic instability 
in accretion disk theory.
Questions to be addressed are abundant, for example 
what is the influence of the aspect ratio $H/R$ of the disk
on the instability? What are the critical values for $\beta_K$?
Can one measure the growth rates for idealized perturbations
with a single wave number $m$?

Apparently  the dissipation and conservation properties of different numerical
hydro schemes affect the development of the instability. We already noticed
that not all available codes show the same results if one does baroclinic
simulations. We will continue with these tests and suggest that other
owners of hydro codes should try to do so as well, in order to prove or disprove our findings.

The next step in our investigations of this new instability will be a 
detailed stability analysis, which has to be the ultimate proof for
our numerical findings.

\acknowledgments
We want to thank 
Steve Balbus,
Robbins Bell,
J\"urgen Blum,
Axel Brandenburg,
Geoff Bryden,
Andi Burkert,
William Cabot,
Pat Cassen,
Jeff Cuzzi,
Sandy Davies,
Steve Desch,
Christian Fendt,
Charles Gammie,
Patrick Godon,
John Hawley,
Thomas Henning,
Sascha Kempf,
Willy Kley,
Greg Laughlin,
Doug Lin,
Mordecai-Mark Mac Low,
Andy Nelson,
Laura Nett,
G\"unther R\"udiger,
Michal R\'o\.zyczka,
Scott Seagroves,
Frank Shu,
Jim Stone,
Paolo Tanga,
Caroline Terquem,
David Trilling,
Dotty Woolum, and 
Richard Young
for support, fruitful criticism and inspirational discussions. 
The results  
are based on calculations
performed at the HLRZ (Hochleistungsrechenzentrum) in  J\"ulich, Germany 
and on the NPACI grant CSC217 and CSC222 in San Diego. 
This research
has been supported in part by the NSF through grant AST-9618548,
by NASA through grant NAG5-4610,  and by a
special NASA astrophysics theory program which supports a joint
Center for Star Formation Studies at NASA-Ames Research Center, UC
Berkeley, and UC Santa Cruz.

\clearpage




\begin{table}%
\begin{tabular}{l|l|r|r|r|l|l|l|l}
 {name} & {grid} & {$R_{i,o}[{ AU}]$}  &  
{$\delta\varphi$} & {$\beta_K$} & {$\alpha = A_{r\phi}$} & {$M$} &  
{$t_{int}/t_{ dyn}$}   \\
\hline
{Model 1} &$51\times20\times60$&3.5\,--\,6.5&90$\arcdeg$&--&$2\times 10^{-3}$ & $0.5$&68\\
{Model 1B} &$51\times20\times60$&3.5\,--\,6.5&90$\arcdeg$&--&$2\times 10^{-2}$ & $1.5$&14\\
{Model 2} &64$^2$    &4.0\,--\,6.0 &30$\arcdeg$ &  0.0   &$2\times 10^{-9}$&$10^{-4}$&110\\
{Model 2B} &64$^2$    &4.0\,--\,6.0 &30$\arcdeg$ &  0.0   &$3\times 10^{-4}$&$0.04  $&1\\
{Model 3} &64$^2$    &4.0\,--\,6.0 &30$\arcdeg$ &  0.57  &$1.5\times 10^{-4}$ &$ 0.03   $&110\\
{Model 4} &128$^2$   &4.0\,--\,6.0 &30$\arcdeg$ &  0.57  &$2\times 10^{-4}$ &$ 0.04    $&130\\
{Model 5} &128$^2$   &3.0\,--\,7.0 &60$\arcdeg$ &  0.57  &$2.5\times 10^{-4}$ &$ 0.05   $&63\\
{Model 6} &128$^2$   &1.0\,--\,10.0&360$\arcdeg$ &  0.57  &$8\times 10^{-3}$ &$ 0.3    $&50
\end{tabular}
\caption{\label{tab1} Parameters chosen in the different
simulations. These parameters are the dimensioning of the grid ($n_r, n_\vartheta,n_\varphi$) or 
($n_r, n_\varphi$),
 the radial extent  of the disk  $R_{i}$  to $R_{o}$,
the azimuthal extent {$\delta\varphi$}, the parameter $\beta_K$
 which describes the variation of
the polytropic $K$ with radius, 
the strength of  the measured Reynolds stresses $\alpha = A_{r\phi}$,
the Mach number ($M$) of the turbulence, and the  ratio of the time
over which the averaging was performed to the orbital period at the
outer edge of the disk. }
\end{table}
\clearpage


\figcaption[fig1.eps]{\label{fig1.ref}Three-dimensional rendering of the
vertical mass flux in model 1, at a time near the end of the 
run.  Red denotes positive velocities,
and blue denotes negative. Compare this figure to Fig.\ 3 in Stone \& Balbus (1996).}

\figcaption[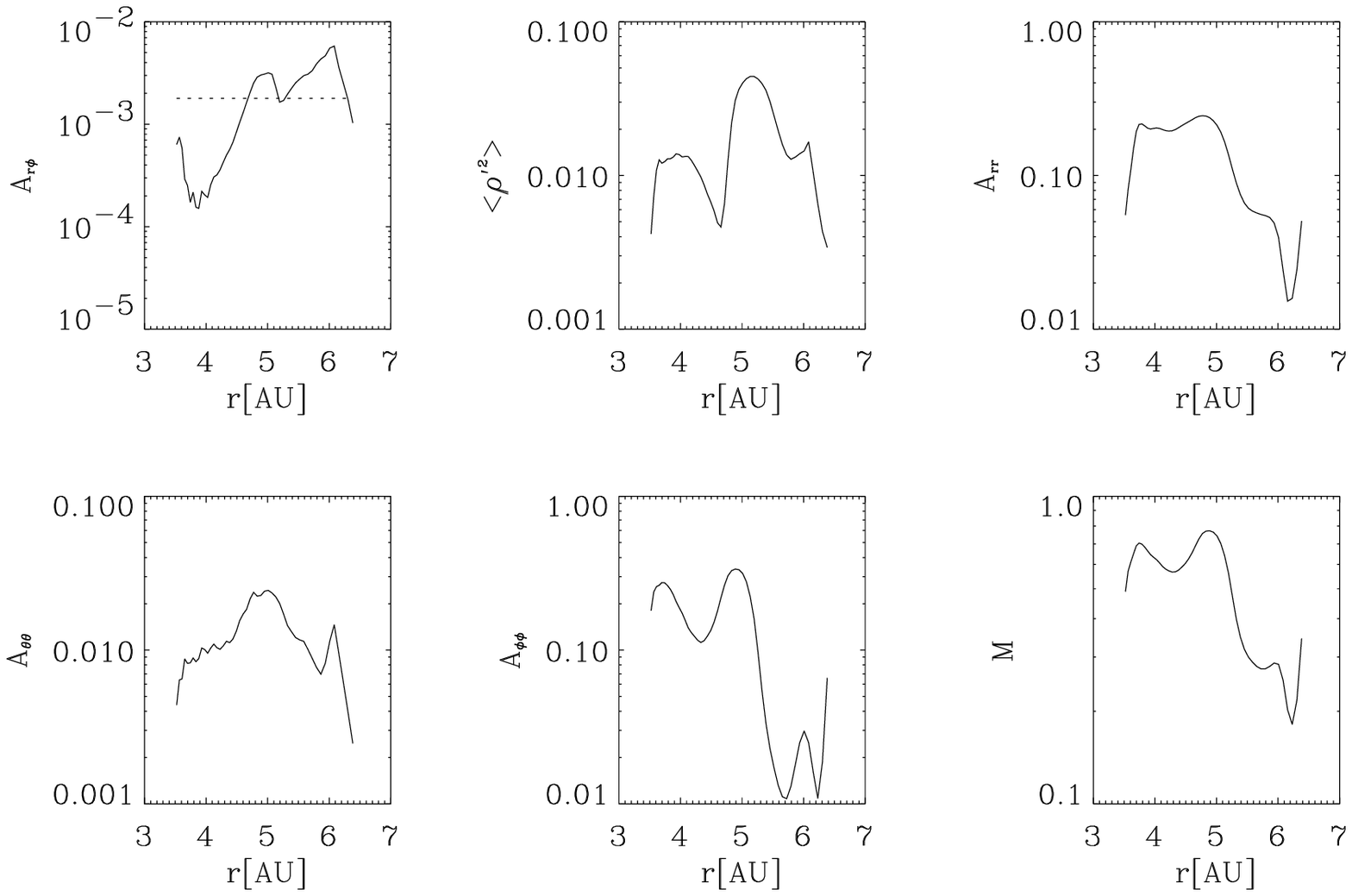]{\label{fig2.ref} Turbulence in model 1: Vertically-, 
azimuthally- 
and time-averaged stresses over 68 orbits, 
measured in units of an effective $A$ (see eq.[ \ref{alpha.ref}])
 and plotted as a function of radius (in AU). 
The mean value for the angular momentum transport ({\it upper left}: $\alpha = A_{r\phi} 
\approx  2 \times 10^{-3}$) is given by the dotted line. Other frames 
give the averages of the relative density fluctuation,
the strength of the turbulence in terms of velocity fluctuations
(eqs.[\ref{alpha_rr.ref}], [\ref{alpha_tt.ref}], and [\ref{alpha_pp.ref}]),
  and Mach number.
See \S 3.2 for explanation.}

\figcaption[fig3.eps]
{\label{fig3.ref} Model 1 at a time near the end of the 
run: Surface density (
colors: $96$ [violet] to $335$ [red] g/cm$^2$),
velocities (vectors:  $v_{max} = 0.6 \times$ the local sound speed) and iso-temperature contours in the midplane.}

\figcaption[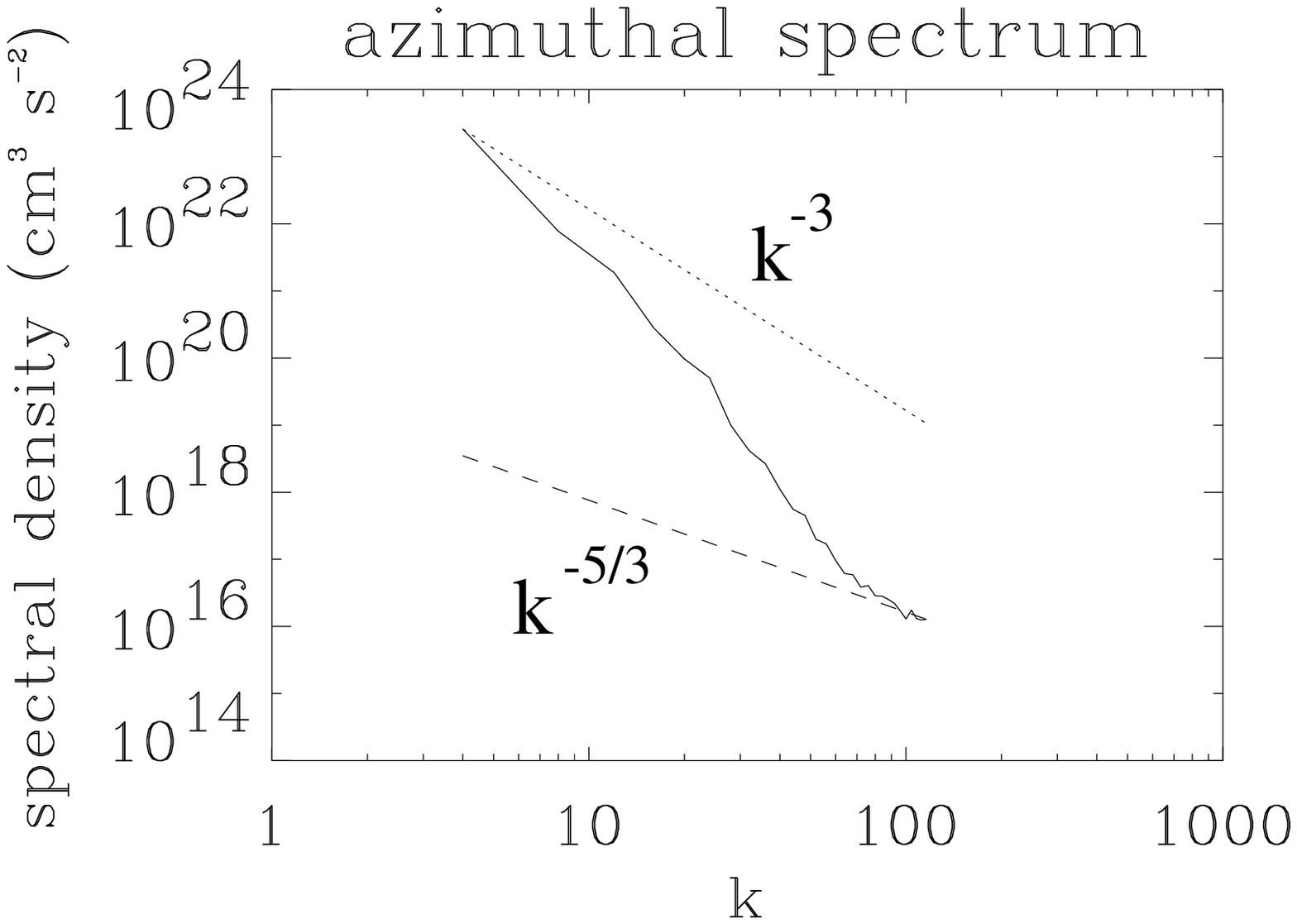]
{\label{fig4.ref} Model 1 at the same time as 
in Fig.\ \ref{fig3.ref}:
Spectral density distribution of the velocities at the midplane computed along the {$\varphi$}-direction and averaged over radius. The slope for isotropic, incompressible turbulence (i.e. a Kolmogorov spectrum) is indicated by the dashed line and
the spectrum for 2D geostrophic flows by the dotted line.}

\figcaption[fig5.eps]
{\label{fig5.ref} Model 1B: Surface density 52 orbits after the initial state (
colors: $20$ [violet] to $523$ [red] g/cm$^2$),
velocities (vectors:  $v_{max} = 0.75 \times$ the local sound speed) and iso-temperature contours in the midplane.}

\figcaption[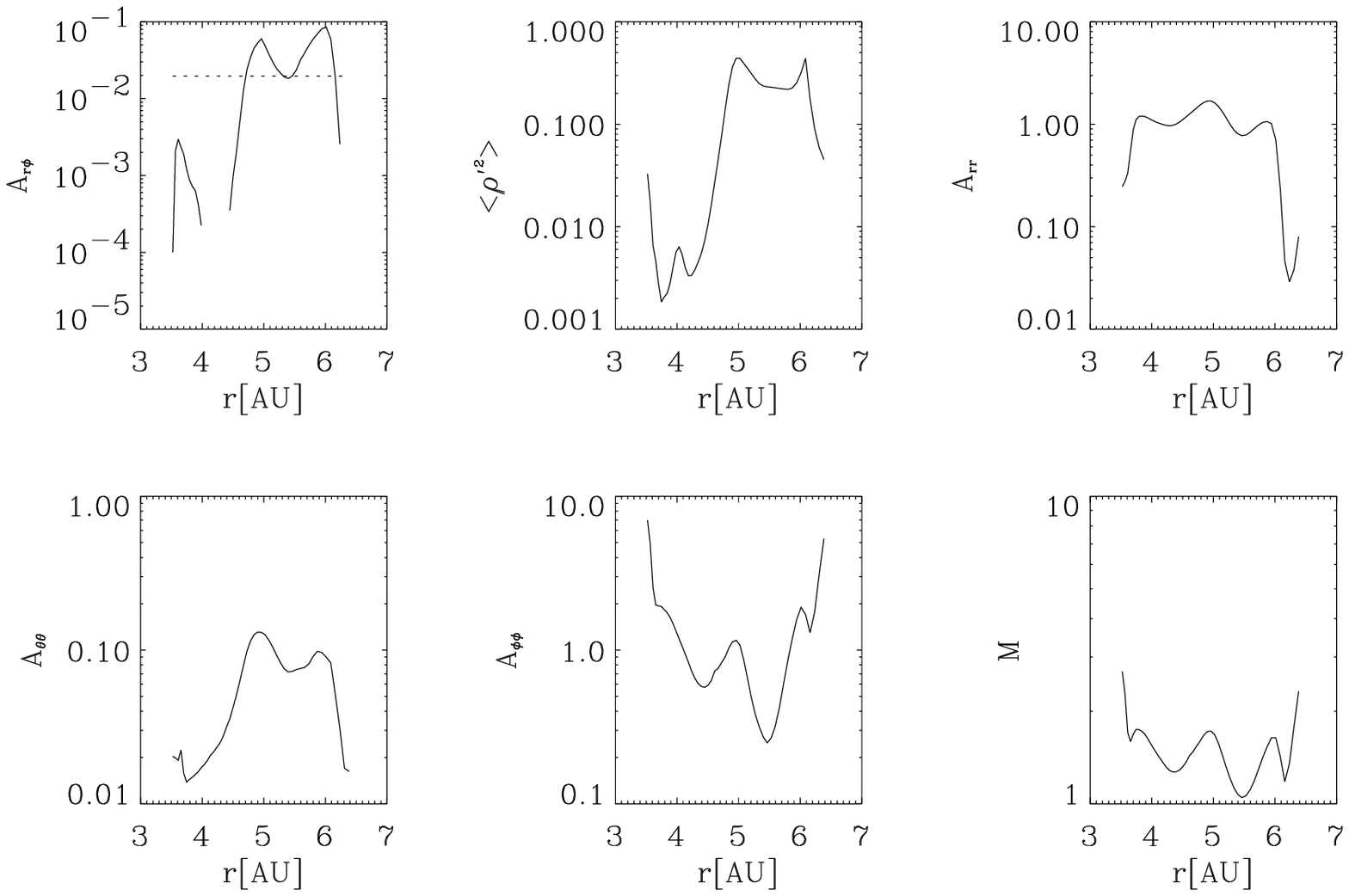]
{\label{fig6.ref} Turbulence in model 1B: Vertically-, 
azimuthally- 
and time-averaged stresses over 14 orbits, 
taken 52 orbits after the initial perturbation,
measured in units of an effective $A$ (see eq.[ \ref{alpha.ref}])
 and plotted as a function of radius (in AU). 
The mean value for the angular momentum transport ({\it upper left}: $\alpha=A_{r\phi} 
\approx  2 \times 10^{-2}$) is given by the dotted line. Other frames 
give the averages of the relative density fluctuation,
the strength of the turbulence in terms of velocity fluctuations
(eqs.[\ref{alpha_rr.ref}], [\ref{alpha_tt.ref}], and [\ref{alpha_pp.ref}]),
  and Mach number.
See \S 3.2 for explanation.} 

\figcaption[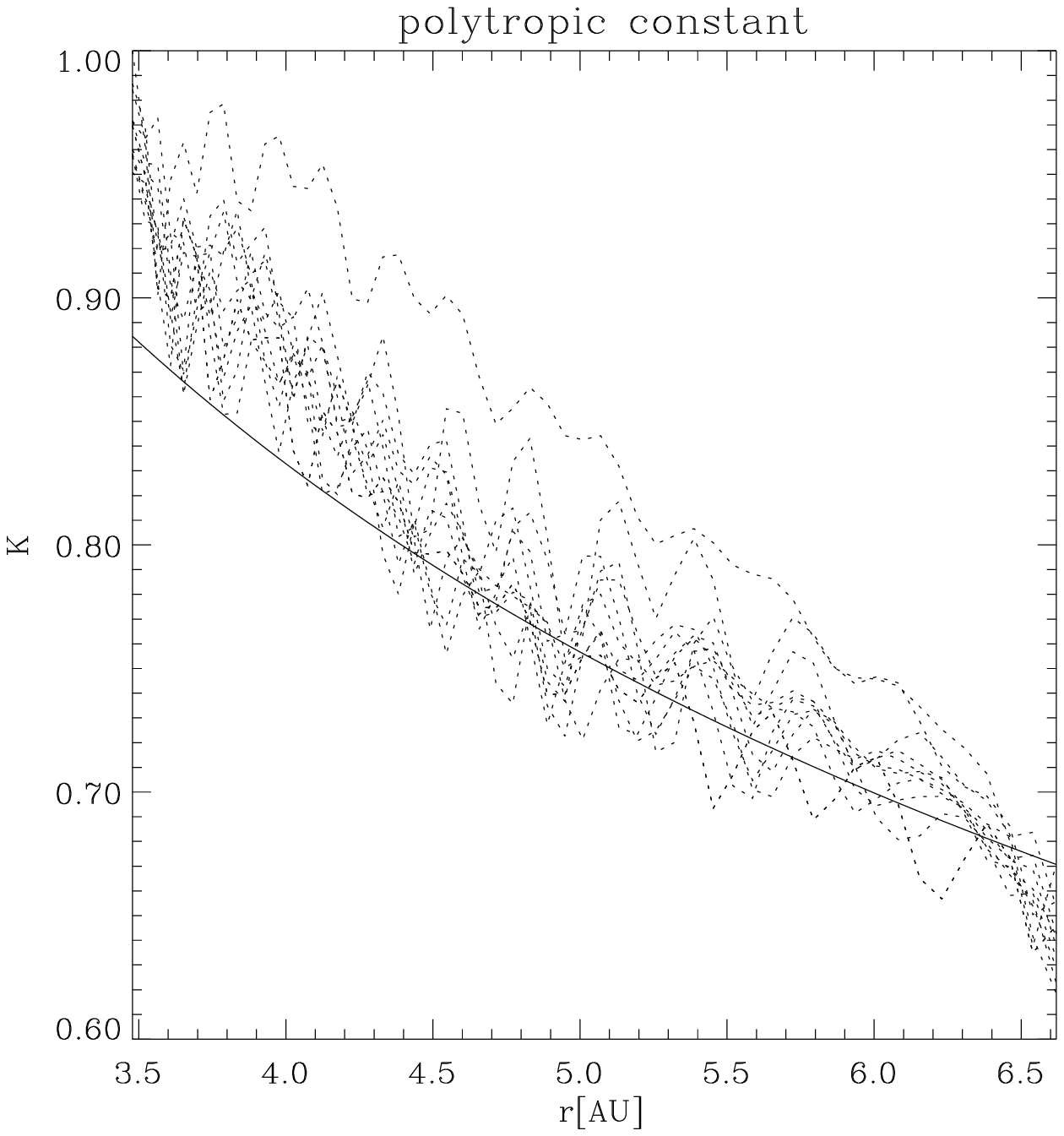]
{\label{fig7.ref} Model 1: Radial
distribution of the polytropic $K$ (normalized to the biggest value) 
as measured from the temperature
and density in the midplane in the thermal convective flow before the system 
became turbulent, i.e.\ it was axisymmetric ({\it dotted lines}: snapshots at 10 different times; {\it solid line}:
theoretical value for $\beta_{K} = 0.57$).}

\figcaption[fig8.eps]
{\label{fig8.ref} Model 2: Surface density (colors: $277$
[violet], $300$ [green]
 to $305$ [red] g/cm$^2$ [same color coding as in Fig.\ \ref {fig12.ref}])
after 90 orbits,
velocities (vectors: $v_{max} = 1 \times 10^{-4} \times$ sound speed) 
and iso-temperature contours in the midplane. 
Barotropic simulation with  no growing non-axisymmetric instability.}

\figcaption[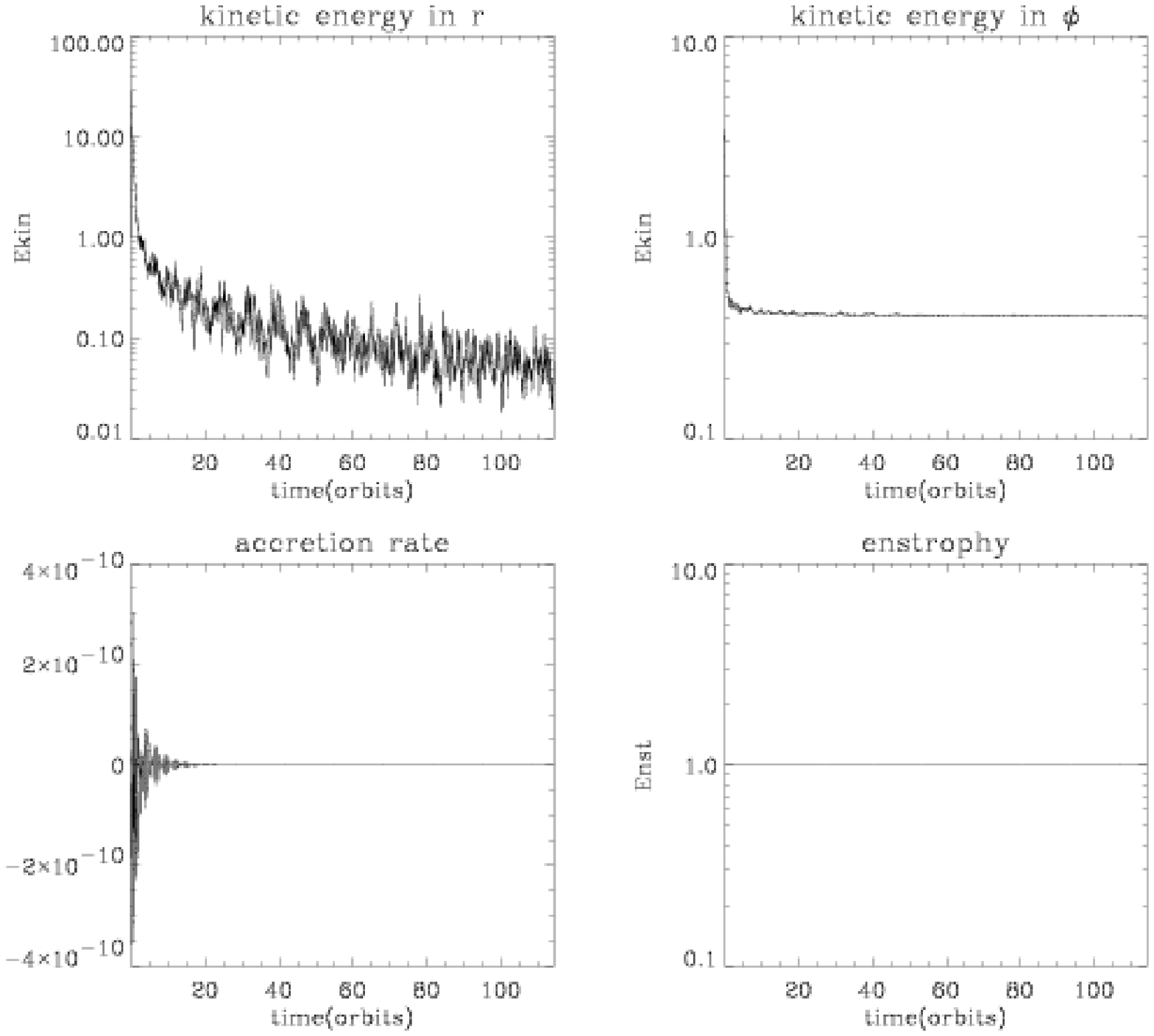]
{\label{fig9.ref} Model 2: Time development of 
kinetic energy in the r-direction ({\it upper left}), in the $\phi$-direction 
({\it upper right}), spatially mean accretion rate in 
M$_\odot$ yr$^{-1}$ averaged over five orbits  ({\it lower  left}), 
and enstrophy (integral square vorticity; {\it lower  right}).  The units for 
kinetic energy and enstrophy are normalized to the first value occurring 
in the data set. As predicted for barotropic
flows, no instability growth can be observed within the first 100 orbits.
No vorticity is generated. 
}

\figcaption[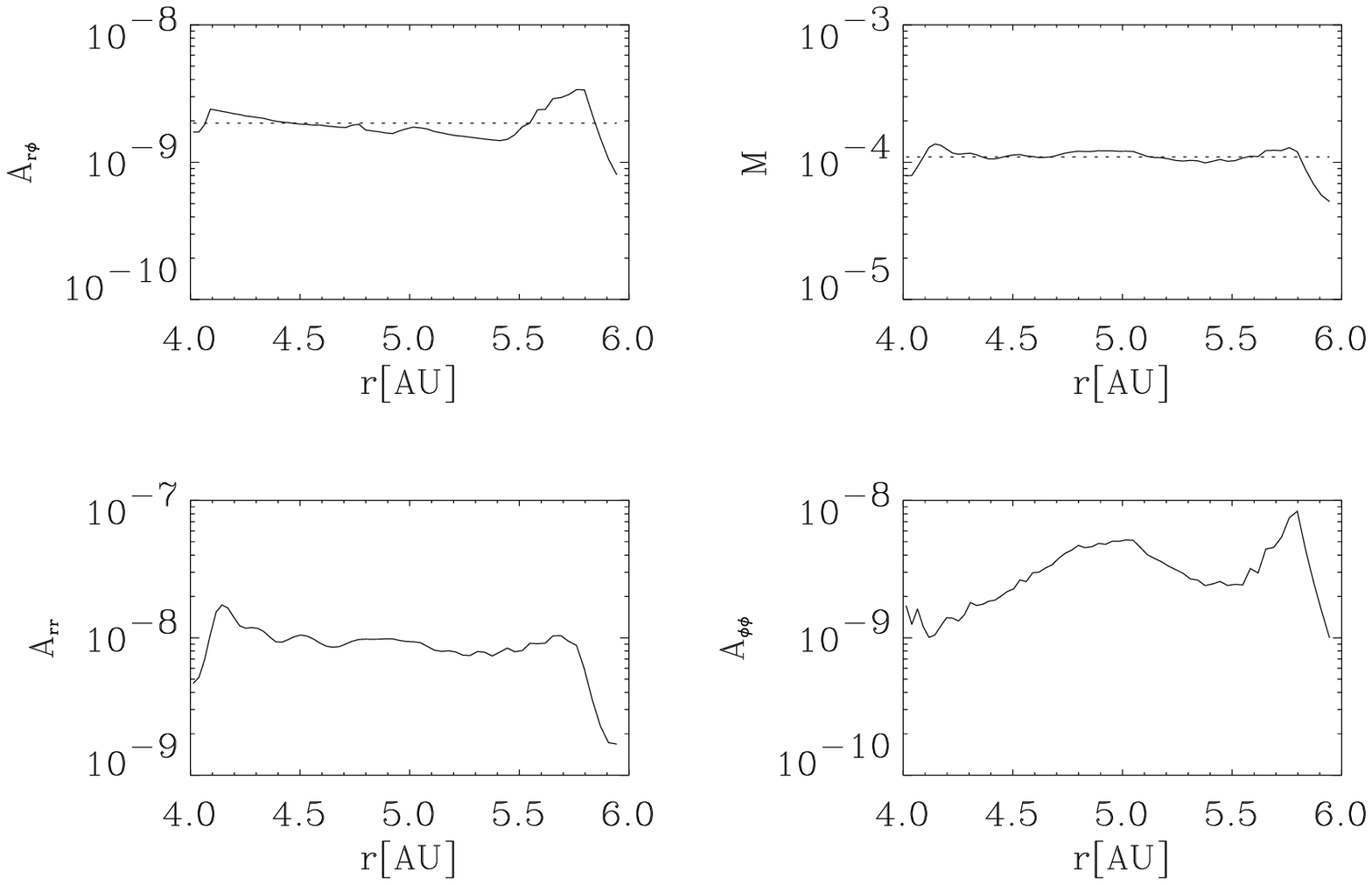]
{\label{fig10.ref} Model 2: Azimuthally- 
and time-averaged Reynolds stress over 10 orbits ({\it upper left})
taken 100 orbits after the initial perturbation, 
measured in units of an effective $A_{r\phi}$ (see eq. [\ref{alpha.ref}]) and plotted 
as a function of radius (in AU). 
The mean value is $\alpha=A_{r\phi} \approx  2.0 \times 10^{-9}$, which is practically zero.
Other frames give the overall Mach number,           
and the  strength of the turbulence in the radial direction ($A_{rr}$; 
see eq. [\ref{alpha_rr.ref}]) and
the azimuthal direction ($A_{\phi\phi}$; see eq. [\ref{alpha_pp.ref}]).}

\figcaption[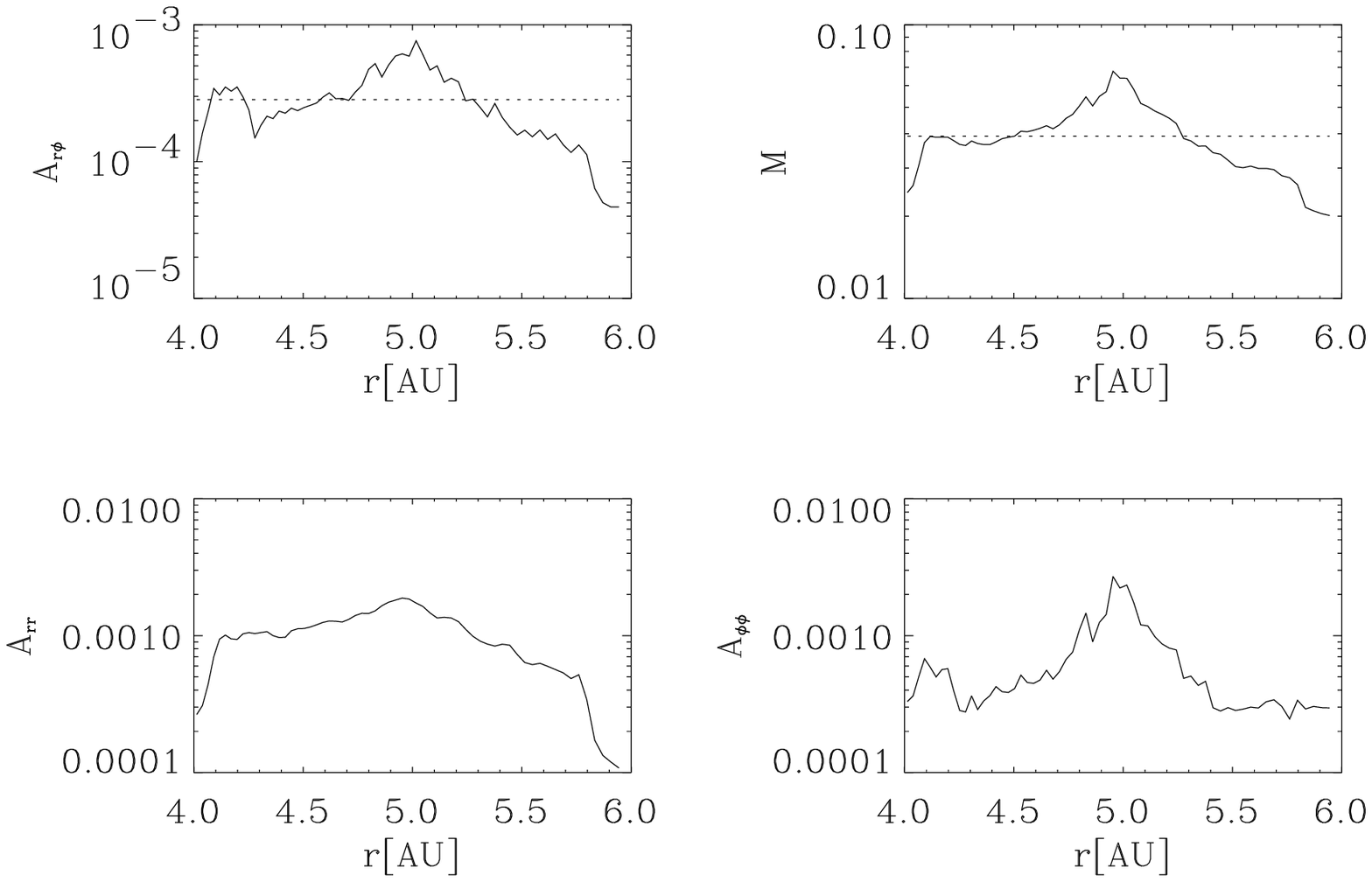]
{\label{fig11.ref} Model 2B: Azimuthally- 
and time-averaged Reynolds stress over 1 orbit ({\it upper left})
right after a strong initial density perturbation, 
measured in units of an effective $A_{r\phi}$ (see eq. [\ref{alpha.ref}]) and plotted 
as a function of radius (in AU). 
The mean value is $\alpha=A_{r\phi} \approx  3 \times 10^{-4}$.
Other frames give the overall Mach number,           
and the  strength of the turbulence in the radial direction ($A_{rr}$; 
see eq. [\ref{alpha_rr.ref}]) and
the azimuthal direction ($A_{\phi\phi}$; see eq. [\ref{alpha_pp.ref}]).}

\figcaption[fig12.eps]
{\label{fig12.ref} Model 3: Surface density (colors: $286$
[violet] to $318$ [red] g/cm$^2$),
velocities (vectors: $v_{max} =  0.03 \times$ sound speed) and iso-temperature contours in the midplane after 600 orbits.
Baroclinic simulation with a still growing non-axisymmetric instability.}

\figcaption[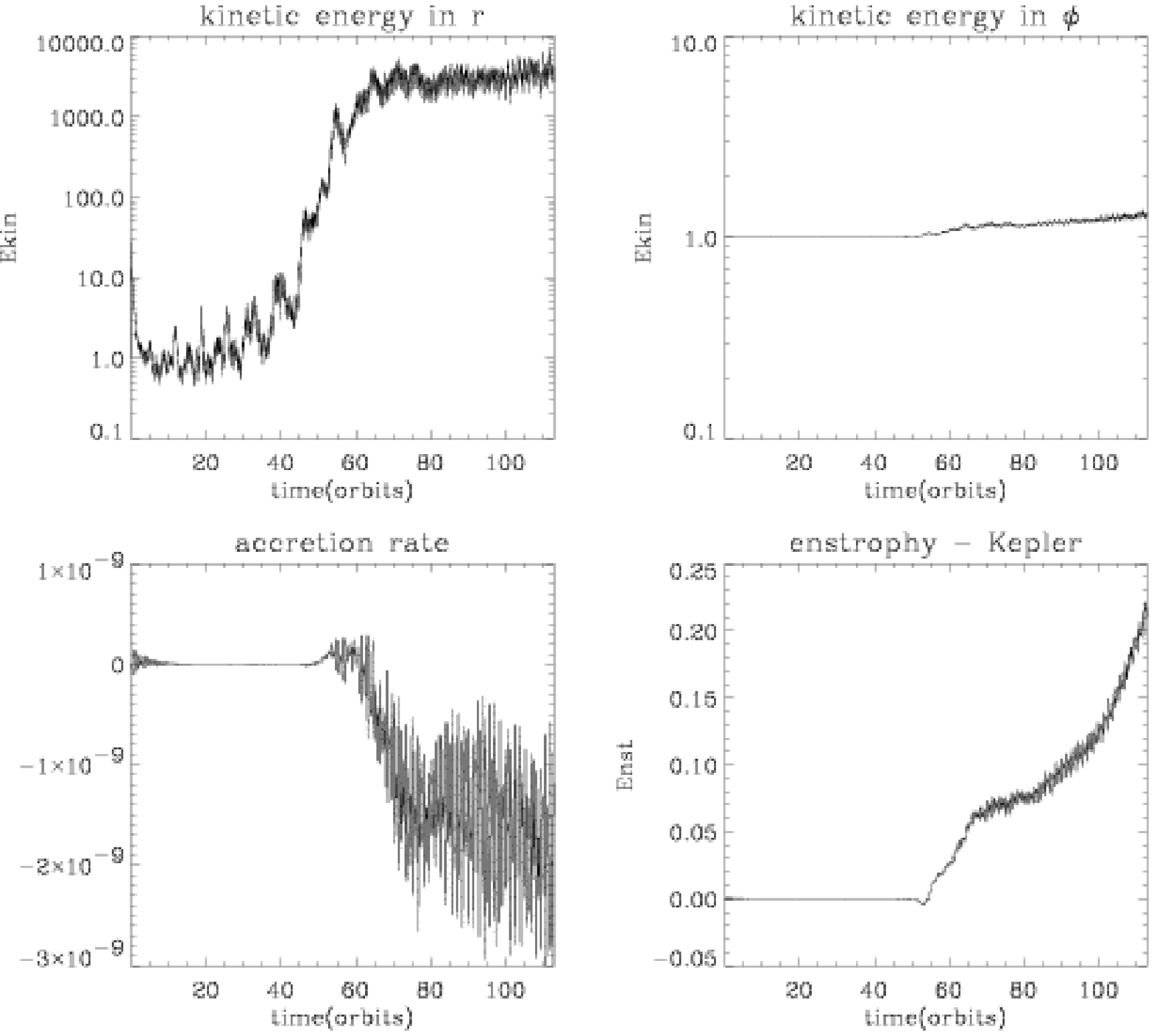]
{\label{fig13.ref} Model 3: The quantities plotted have the same
meaning as those in Fig.\ \ref{fig9.ref} and are directly comparable.
In baroclinic
flows, vorticity and enstrophy are created and instability 
growth can be observed within the first 100 orbits.
}

\figcaption[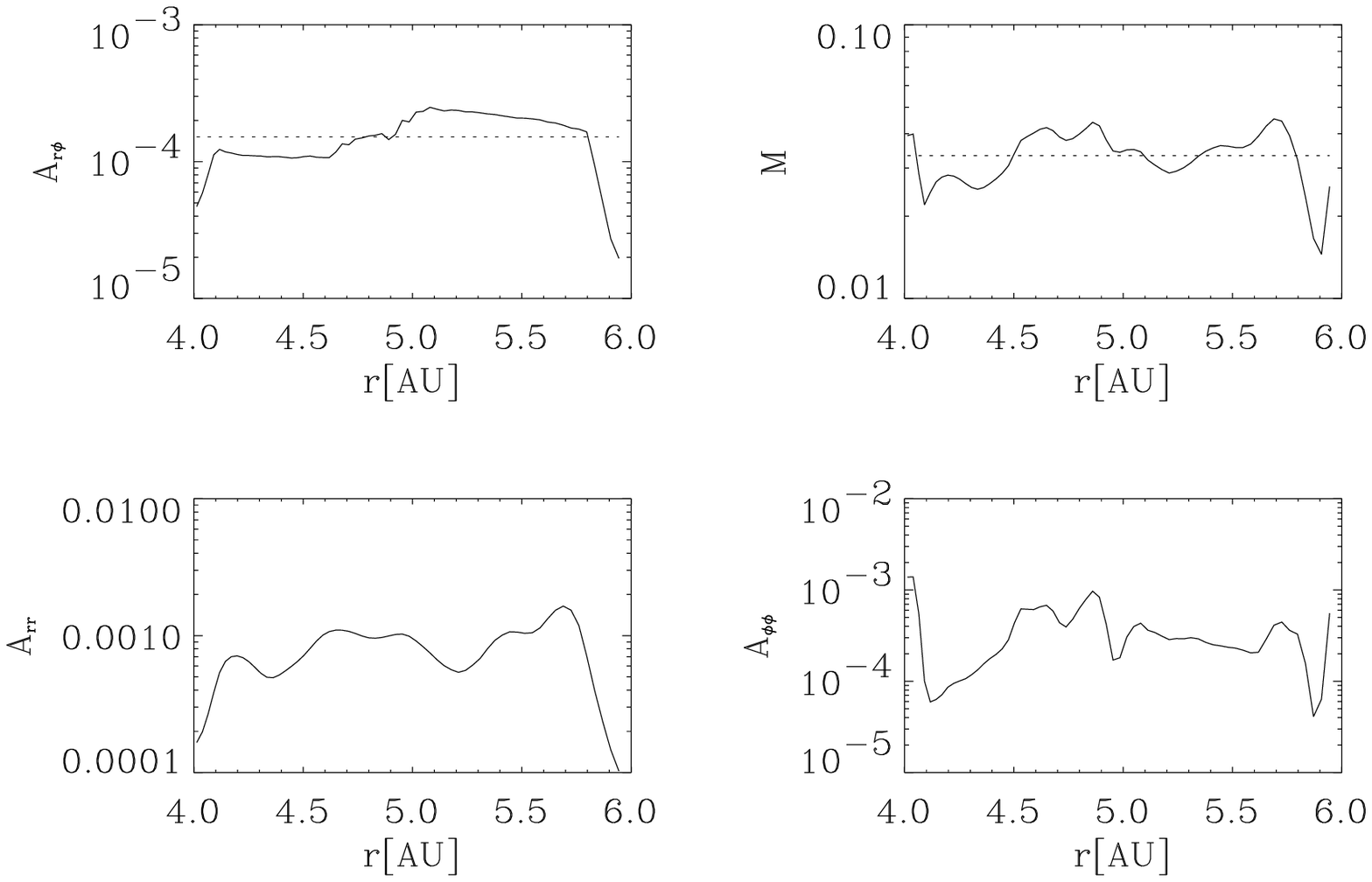]
{\label{fig14.ref} Model 3:  Azimuthally- 
and time-averaged Reynolds stress over 380 orbits (starting with 2030 orbits after the initial perturbation),
measured in units of an effective $A_{r\phi}$ (see eq. [\ref{alpha.ref}]) and plotted 
as a function of radius (in AU). 
The mean value ({\it upper left}) is $\alpha=A_{r\phi} \approx  1.5 \times 10^{-4}$. 
 Other frames show the overall Mach number and 
the strength of the turbulence in the radial direction
($A_{rr}$; see eq. [\ref{alpha_rr.ref}])
 and in the  azimuthal direction ($A_{\phi\phi}$; see eq. [\ref{alpha_pp.ref}]).}

\figcaption[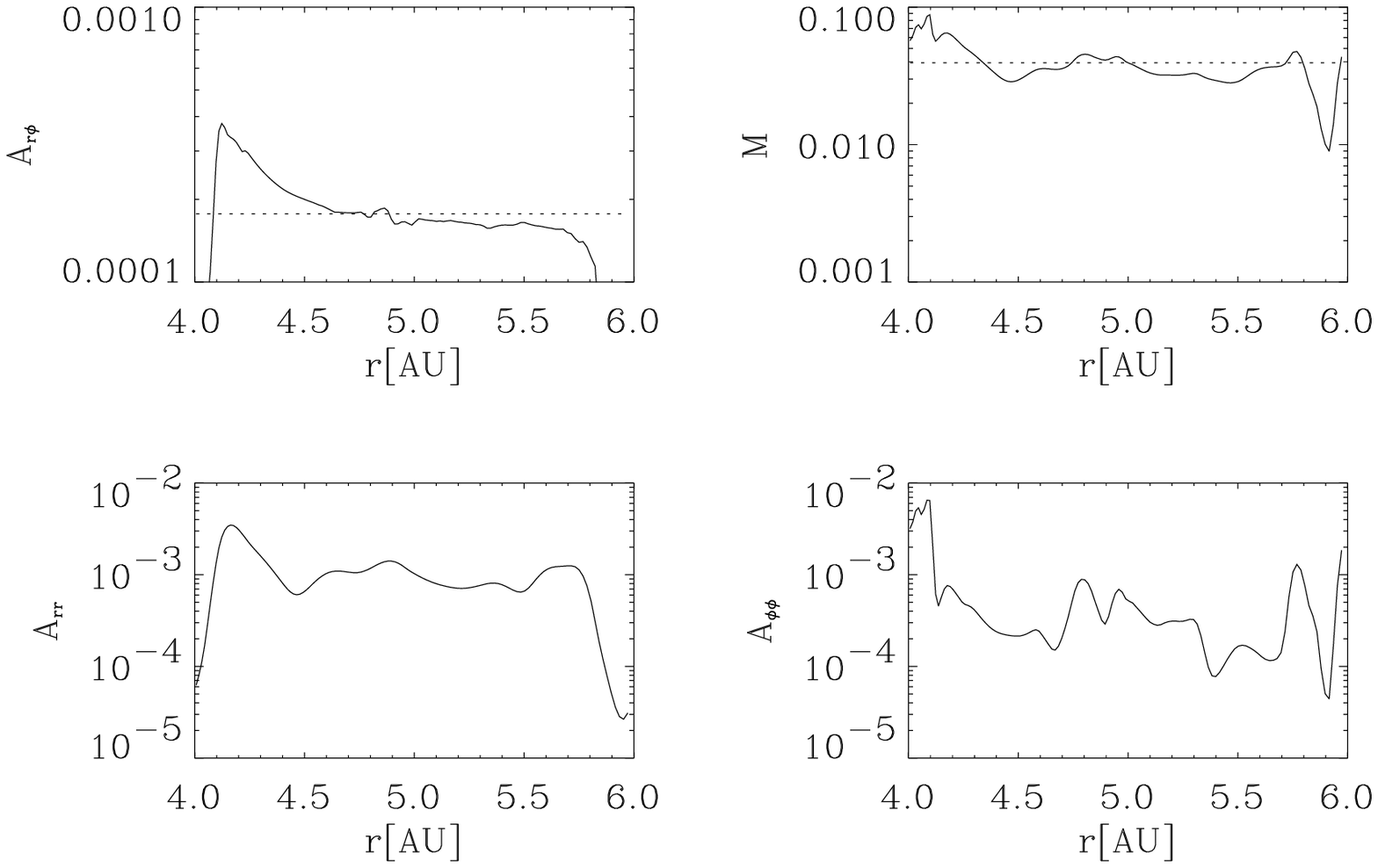]
{\label{fig15.ref} Model 4: Stresses averaged 
over 380 orbits (starting with 2030 orbits after the initial perturbation).
The quantities plotted have the same meaning
as in Fig.\ \ref{fig14.ref}.
The mean value of $\alpha=A_{r\phi} \approx 2\times 10^{-4}$.}

\figcaption[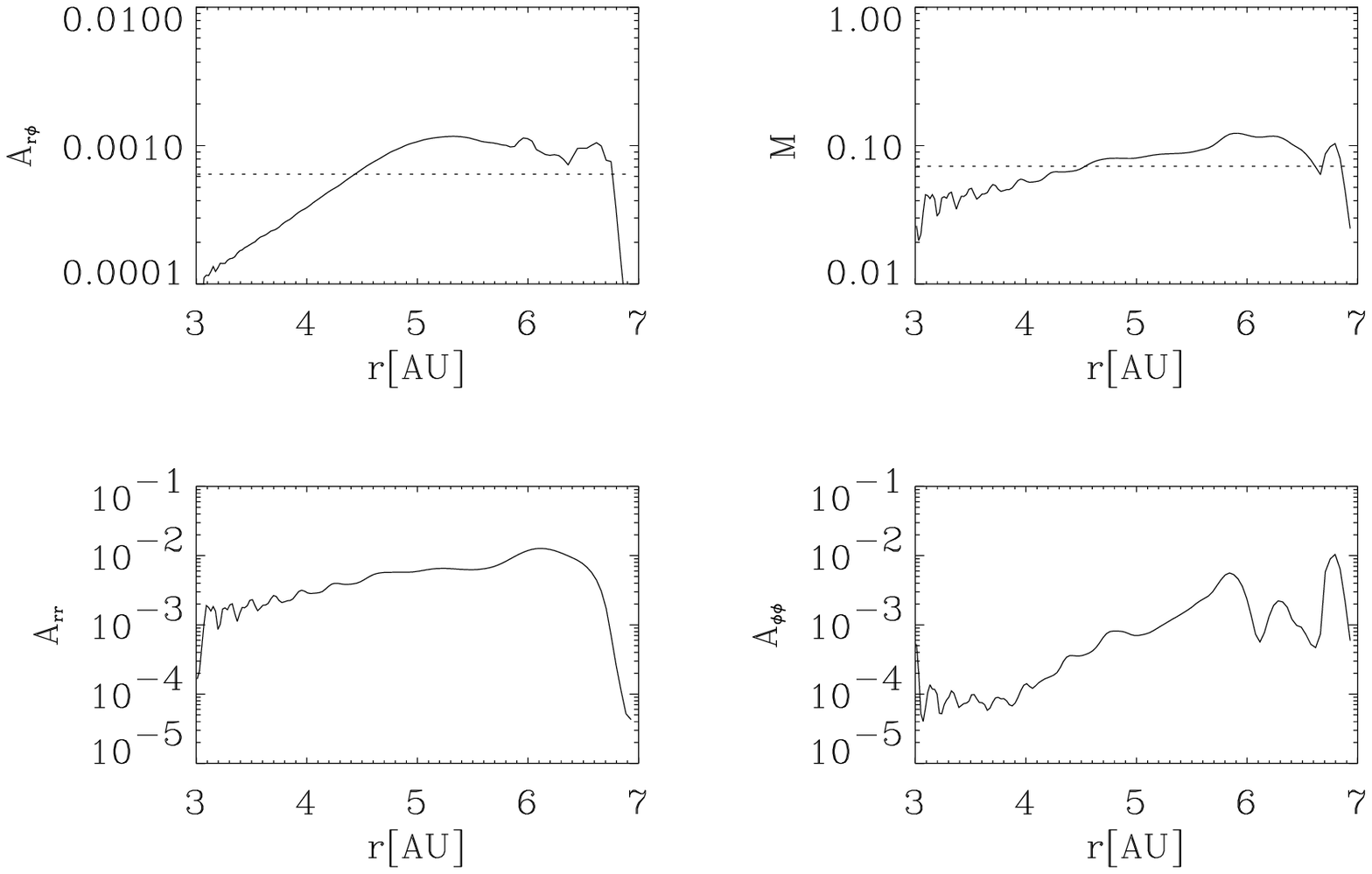]
{\label{fig16.ref} Model 5: Stresses averaged 
 over 180 orbits (starting with 600 orbits after the initial perturbation).
The quantities plotted have the same
meaning as in Fig.\ \ref{fig14.ref}. 
The mean value of $\alpha=A_{r\phi} \approx  5.0 \times 10^{-3}$.}

\figcaption[fig17.eps]
{\label{fig17.ref} Model 5: The quantities plotted have the same
meaning as those in Fig.\ \ref{fig12.ref}.
This calculation was run over 230 orbits.}

\figcaption[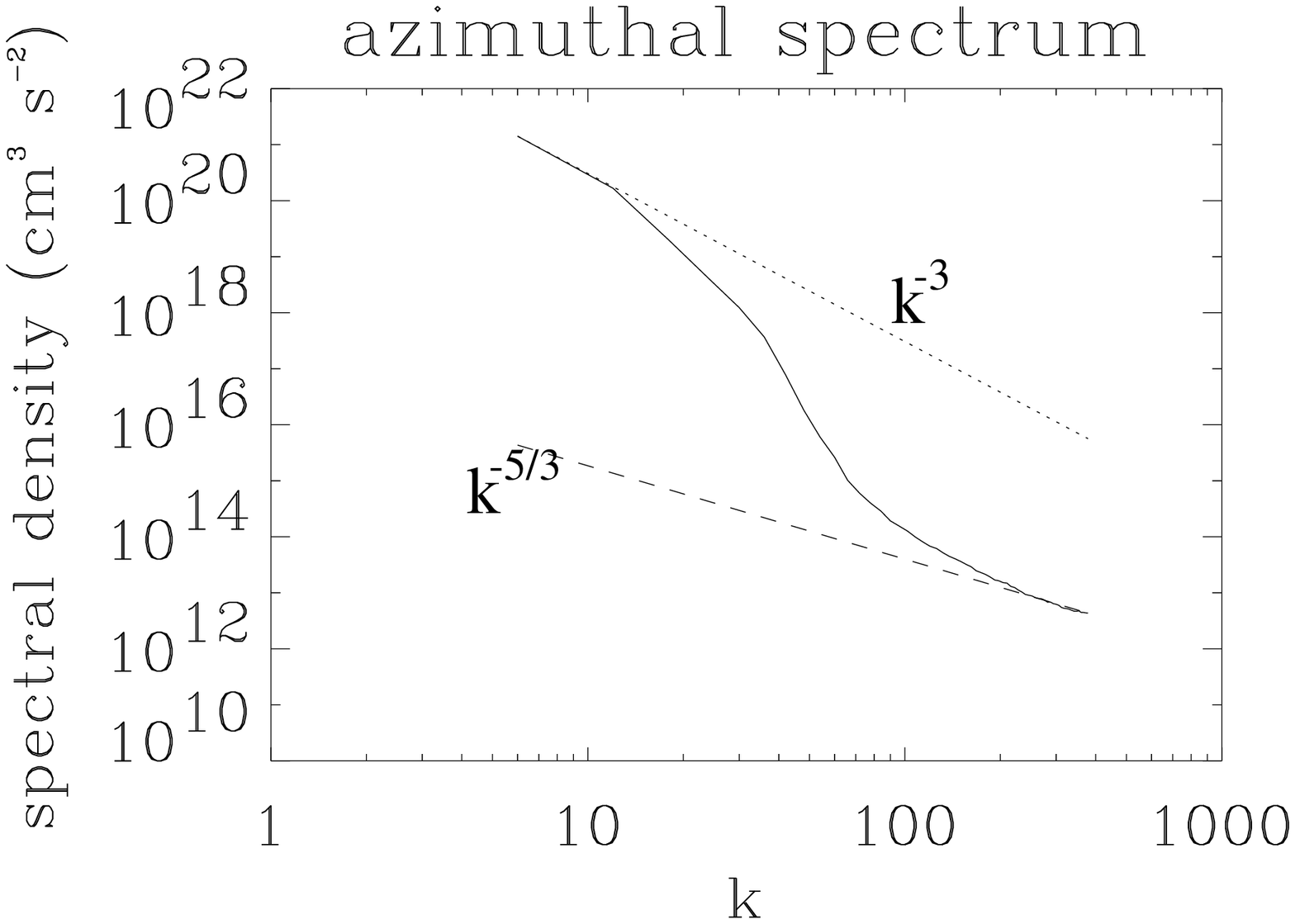]
{\label{fig18.ref} Model 5: Spectral density distribution of the velocities at the midplane computed along the {$\varphi$}-direction and averaged over radius. The slope for isotropic, incompressible turbulence (i.e. a Kolmogorov spectrum) is indicated by the dashed line and
the spectrum for 2D geostrophic flows by the dotted line.}

\figcaption[fig19.eps]
{\label{fig19.ref} Model 6: Four snapshots of the evolution of surface density (colors: $650$ [red], $550$ [yellow], $450$ [green], $250$ [blue] to  $<100$ [black] to g/cm$^2$)
in the global model in polar ($r-\phi$) coordinates. 
The times  are  a): 1, b): 95, c): 190 and d): 320 orbits at the outer radius. Contours
of equal pressure are also shown. }

\figcaption[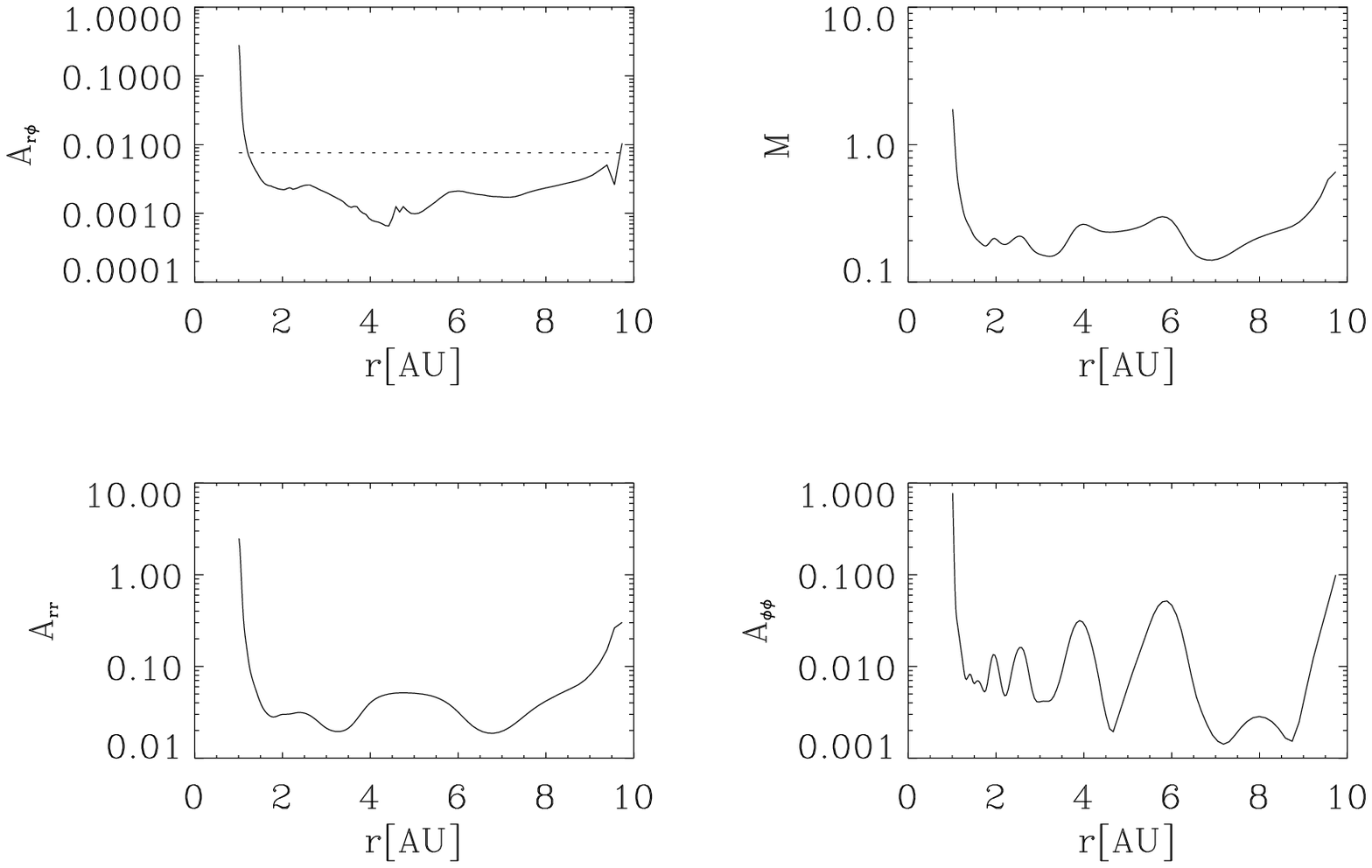]
{\label{fig20.ref} Model 6: Stresses averaged 
 over 50 orbits (starting with 190 orbits after the initial perturbation).  Quantities plotted have the same meaning 
as in Fig.\ \ref{fig10.ref}.  The mean value of $A_{r\phi} \approx  8.0 \times 10^{-3}$.}

\figcaption[fig21.eps]
{\label{fig21.ref} The ``pre-protoplanet'' in Model 6:
 Surface density (colors: $650$ [red], $550$ [yellow], $450$ [green], $250$ [blue] to  $<100$ [black] to g/cm$^2$) in the global model is projected in a cartesian frame after 
320 orbits at the outer radius, which corresponds to $10^{4}$ yr.
 Note that the condensation  is partially artificially smeared
out in the $\varphi$-direction, which is a result of the low order advection scheme.
 In reality one could expect the ``pre-protoplanet'' to be more strongly confined.}

\figcaption[fig22.eps]
{\label{fig22.ref} Model 6: Surface density (colors: $650$ [red], $550$ [yellow], $450$ [green], $250$ [blue] to  $<100$ [black] to g/cm$^2$) and velocity (vectors:  $v_{max} \approx 1.5 \times$ sound speed)
 in the global model in polar ($r-\varphi$) coordinates after 
320 orbits at the outer radius, which corresponds to $10^{4}$ yr. 
Plotted velocities  in the $\varphi$ direction 
are obtained by subtracting the mean azimuthal velocity at each radius, which explains
why the vortex center seems to be displaced from the density maximum. }

\end{document}